%% file: main.tex
\documentclass[twocolumn]{article}
\usepackage[utf8]{inputenc}
\usepackage[english]{babel}
\usepackage{amsfonts}
\usepackage{amsmath,amssymb,amsthm}
\usepackage{physics}
\usepackage{mathdots}
\usepackage{mathtools}
\usepackage{graphicx}
\usepackage{fancyhdr}
\usepackage{lastpage}
\usepackage{changepage}
\usepackage{lipsum}
\usepackage{pgfplots}
\usepackage{bbm}
\usepackage{fancyhdr}
\usepackage{listings}
\usepackage{comment}
\usepackage{mathrsfs}
\usepackage{blindtext}
\usepackage{enumitem}
\usepackage{xcolor}
\usepackage{placeins}
\usepackage{ulem}
\usepackage{longtable}
\usepackage{booktabs}      
\usepackage{caption}       
\usepackage{float}
\usepackage{stfloats}      
\usepackage{afterpage}
\usepackage{adjustbox}
\usepackage{cuted}         
\usepackage{blindtext}
\usepackage[T1]{fontenc}
\usepackage{hyperref}
\usepackage[numbers,sort&compress]{natbib}
\hypersetup{
    colorlinks=true,
    linkcolor=blue,
    filecolor=magenta,      
    urlcolor=cyan,
}
\usepackage{graphicx,wrapfig,lipsum}
\usepackage{xcolor}
\usepackage{subcaption}
\definecolor{jcol}{RGB}{0,128,128} 
\title{How do standard time series forecasting benchmarks translate to finance? A broad study of risk adjusted performance\footnote{Git Repository \href{https://github.com/Adir7SK/deep_fin_benchmark}{here}.}}

\author{
Adir Saly-Kaufmann$^{1}$,
Kieran Wood$^{1,2}$,
Jan-Peter Calliess$^{1}$,
Stefan Zohren$^{1}$ \\
\\
$^{1}$Machine Learning Research Group, Department of Engineering Science, University of Oxford \\
$^{2}$Oxford-Man Institute of Quantitative Finance, University of Oxford \\
\\
\texttt{adir.saly-kaufmann@eng.ox.ac.uk, kieran.wood@eng.ox.ac.uk} \\
\texttt{jan@robots.ox.ac.uk, stefan.zohren@eng.ox.ac.uk}\\
}
\date{}   

\begin{document}

\maketitle

\begin{abstract}
    \label{sec:abstract}
    \input{Sections/abstract.tex}
\end{abstract}

\section{Introduction}
\label{sec:introduction}
\input{Sections/introduction.tex}

\section{Architectures}
\label{sec:architectures}
\input{Sections/architectures.tex}


\section{Empirical Results}
\label{sec:benchamrk}
\input{Sections/benchmark.tex}

\section{Conclusions}
\label{sec:conclusion}
\input{Sections/conclusion.tex}



\bibliographystyle{unsrtnat}
\bibliography{references}

\clearpage
\appendix
\label{sec:appendix}
\input{Sections/appendix.tex}

\end{document}

%% file: Sections/abstract.tex

Forecasting financial asset returns is challenging due to low signal-to-noise ratios, non-stationarity and heavy-tailed distributions. We benchmark 16 deep learning architectures for time series forecasting, including Transformers, State Space Models (SSMs), xLSTMs and recurrent hybrids, on a 15-year daily futures dataset using an end-to-end volatility-targeted Sharpe ratio optimisation task. Our results challenge prevailing assumptions in the literature. Several widely used 'state-of-the-art' architectures, including generic Transformers and SSMs, underperformed a simple AR(1) benchmark in a realistic financial forecasting task. The strongest results came from comparatively simple hybrid approaches using recurrent structures: a Variable Selection Network combined with an LSTM achieved the highest Sharpe ratio, while xLSTM showed the greatest robustness to transaction costs. These findings suggest that success on standard time-series benchmarks does not reliably transfer to finance, and that architectures tailored to the unique properties of asset returns may remain more effective than generic attention-based approaches for financial forecasting.

%% file: Sections/introduction.tex

\begin{figure}[h]
    \centering
    \includegraphics[width=\columnwidth]{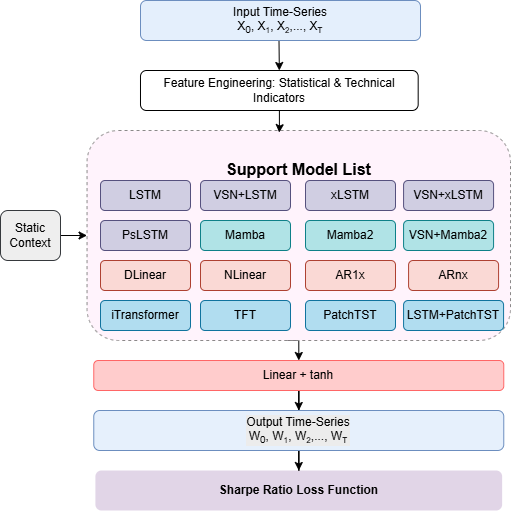}
    \caption{End-to-end portfolio optimization pipeline: Statistical and technical indicators are extracted from historical close prices, serving as the predictive model's inputs. The model outputs are transformed into portfolio weights via a linear projection followed by a hyperbolic tangent activation. Training is performed by minimizing the negative Sharpe Ratio.} 
    \label{fig:implementation_diagram}
    
\end{figure}

In recent years, many new deep learning architectures have emerged in the context of time-series forecasting, with 
Transformer-based architectures \cite{transformer} drawing particular attention \cite{transformer_impact}. 
Several theoretical extensions and adaptations have been proposed, such as iTransformer \cite{itransformer}, PatchTST \cite{patchtst}, and the Temporal Fusion Transformer \cite{tft}. 
Each of these models aims to address different challenges encountered in real-world forecasting. 
For example, PatchTST improves robustness by bundling data points and interpreting features independently, whereas iTransformer focuses on learning relationships between features without relying on temporal order.

State Space Models have also presented an alternative to transformer-based architectures. 
In particular, Mamba2 \cite{mamba2} claims to achieve a mathematically principled architecture with linear-attention-like behavior \cite{lin_attn} while supporting arbitrarily large lookback windows. 
More broadly, State Space Models have evolved from the theory of HiPPO (High-order Polynomial Projection Operators) matrices \cite{hippo}, which maintain compressed representations of past information as the model evolves.

Historically, recurrent neural networks (RNNs) \cite{rnn} have been the most widely used deep-learning architectures for time-series forecasting \cite{rnn_use}, especially LSTMs \cite{lstm}. 
More recently, the xLSTM architecture \cite{xlstm} has been introduced as a new state-of-the-art model, with additional gains demonstrated by PsLSTM \cite{pslstm}. 
xLSTM replaces LSTM’s traditional sigmoid gating with exponential gating and a normalization term, while also employing a memory matrix rather than a scalar value. 
PsLSTM further integrates patching (similar to PatchTST) into the xLSTM architecture.

Several benchmark studies have explored deep-learning models for time-series forecasting \cite{tslib}, focusing on applications such as weather prediction, electricity transformer temperatures, and traffic data. 
These studies indicate that simple architectures such as DLinear and NLinear \cite{nlinear} can perform comparably to more complex transformer-based models.
However, these datasets exhibit strong seasonality and a high signal-to-noise ratio, in contrast to financial time series \cite{fin_ts_characteristics}. 

Forecasting financial time series essentially reduces to performing auto-regressive modeling \cite{ar1}, typically formulated as
\begin{equation}
    x_{t+1} = f(x_{t}, x_{t-1}, x_{t-2}, ...)
    \label{eq:ar_model}
\end{equation}
for each $0\leq t < T$, where $x_t$ is the value of the process at time $t$, $T$ is the last time-step to predict, and $f(\cdot)$ denotes a model-dependent forecasting function that maps past observations to a one-step-ahead prediction. For example, in $AR(p)$ it is of the form:
\[
f(x_t, \dots, x_{t-p}) = \sum_{i=0}^{p} \phi_i x_{t-i},
\]
where $\{\phi_i\}$ are fixed coefficients~\cite{ar1} and in $LSTM$~\cite{lstm}, $f$ is instead represented by a nonlinear recurrent mapping with learnable gating mechanisms and hidden states, see~\ref{appdx:lstm}.

Financial return series are characterized by strong noise, weak and time-varying predictability, and pronounced non-stationarity. These properties imply that predictive success hinges less on raw model capacity than on the ability to recover economically meaningful signals from highly volatile data with low signal-to-noise ratios. This raises the possibility that widely cited performance rankings may not generalise to financial forecasting. Given these distinctive properties, a critical, yet underexamined, question emerges: how well do performance rankings from standard time-series forecasting benchmarks generalise to the low signal-to-noise, highly volatile environments characteristic of financial asset returns where the target (objective) is to maximize the risk-adjusted returns (such as Sharpe Ratio)? In this context, a successful model must simultaneously (i) improve the signal-to-noise ratio by filtering out transient fluctuations, (ii) learn asset-specific dynamics rather than imposing homogeneous temporal structures, and (iii) embed temporal dependencies in a manner that remains stable across market regimes \cite{char_fin_ts1, char_fin_ts2}. Several of the benchmarked architectures incorporate explicit design choices aimed at enhancing one or more of these properties, including feature selection, structured state representations, and temporal aggregation mechanisms \cite{success_fin_ts_models}. We evaluate both the incremental value of such architectural enhancements and their performance relative to models that natively embed these characteristics, with particular emphasis on robustness across time and economic regimes.

This paper provides a systematic empirical answer to the question posed above. Our results challenge prevailing assumptions and offer practical guidance for forecasters navigating the proliferation of deep learning architectures.

\paragraph{Contributions.}
This paper presents a unified benchmark of modern deep learning architectures for financial time-series prediction under a Sharpe-ratio optimisation framework. Using 15 years of data spanning multiple asset classes and market regimes, we evaluate linear models, recurrent networks, transformer-based architectures, state-space models, and recent sequence-representation approaches across return, risk, robustness, and computational dimensions. We are identifying properties that characterise successful models based on empirical observations. To support replication and practical adoption, we release our code as open-source guidelines for implementation.

\medskip

\noindent\textbf{Linear dynamics alone appear insufficient.}
While linear models occasionally performed competitively in specific high-volatility subperiods, they failed to deliver stable performance across time and provided limited incremental value relative to a passive benchmark. This supports the view that financial returns exhibit structural features not fully captured by linear autoregression.

\medskip

\noindent\textbf{Architectural inductive bias is decisive.}
Nonlinear models improved average performance, but outcomes varied substantially across architectures. Generic transformers and state-space models displayed heterogeneous, regime-sensitive behavior. In contrast, VLSTM—designed to learn structured temporal representations—delivered consistently strong and stable risk-adjusted returns, suggesting that representation compression, adaptive memory, and temporal abstraction are particularly valuable in low signal-to-noise environments.

\medskip

\noindent\textbf{Robustness and risk control matter as much as returns.}
Downside exposure, tail behavior, and stability under reduced seed aggregation materially affected model suitability. VLSTM-based strategies combined competitive returns with moderate drawdowns and remained stable under weaker experimental budgets, indicating that performance was not driven solely by favorable seed selection.

\medskip

\noindent\textbf{Asymptotic efficiency does not guarantee empirical superiority.}
Although state-space models offer attractive theoretical complexity, empirical effectiveness depended more strongly on inductive bias than on asymptotic scaling alone.

\medskip

Overall, the results suggest that effective financial forecasting models benefit from jointly denoising returns, learning asset-specific and regime-aware dynamics, and encoding temporal structure in a stable and adaptive manner. While the conclusions remain conditional on the dataset and backtesting protocol employed, this benchmark provides a transparent reference point for future research in deep learning for finance.

%% file: Sections/architectures.tex
\subsection{Problem Setup}

Let $\{x_t\}_{t=1}^T$, $x_t \in \mathbb{R}^d$, denote a multivariate time series of end of day features derived from commodity futures, foreign-exchange (FX) futures, bonds, index, and energy products, including prices, returns, and technical indicators. 

Given a fixed lookback window of length $L$, the objective is to learn a function
\begin{equation}
    f_\theta: \mathbb{R}^{L \times d} \rightarrow [-1, 1],
\end{equation}
mapping historical observations
\begin{equation}
    \mathbf{X}_t = [x_{t-L+1}, \dots, x_t]
\end{equation}
to a scalar forecast or trading signal used to construct daily positions,
where 1 is the upper bound for a full long position and -1 is the lower bound for a full short position. 

\paragraph{Trading Signal Generation} To systematically benchmark different deep learning paradigms, we structure the function $f_\theta$ as a modular, two-stage pipeline. The first stage consists of a candidate sequence architecture $g_\phi$ (e.g., LSTM, PatchTST, or Mamba2), which processes the input window to extract a temporal state representation of fixed hidden dimension $H$:
\begin{equation}
    h_t = g_\phi(\mathbf{X}_t), \quad h_t \in \mathbb{R}^H.
\end{equation}

The second stage is a unified projection head applied to the terminal hidden state $h_t$. This consists of a linear transformation followed by a hyperbolic tangent ($\tanh$) activation function to bound the output:
\begin{equation}
    \hat{y}_t = \tanh(w_{\mathrm{lin}}^\top h_t + b_{\mathrm{lin}}),
\end{equation}
where $w_{\mathrm{lin}} \in \mathbb{R}^H$ and $b_{\mathrm{lin}} \in \mathbb{R}$ are learnable weights. 
All models are trained using rolling windows and evaluated in a fully out-of-sample trading framework. \textbf{We added ticker embeddings to all the models} to enhance the learning per individual ticker/asset~\cite{entity_embeddings}. 
\paragraph{Portfolio Construction}
The scalar output $\hat{y}_{t,k} \in [-1, 1]$ generated by the projection head represents the model's directional conviction for asset $k$ at time $t$. Because financial assets exhibit vastly different baseline volatilities, we employ a volatility targeting framework to equalize risk contributions across the universe~\cite{tsmom,vol_target,enhancing_tsmom}. We estimate the ex-ante conditional volatility $\sigma_{t,k}$ for each asset using an Exponentially Weighted Moving Average (EWMA) estimator (detailed in Appendix \ref{appdx:eda}). This estimation induces a time-varying leverage factor, defined as $\text{vs\_factor}_{t,k} = \frac{1}{\sigma_{t,k}}$, which dynamically scales position sizes in response to shifting market regimes.
Given a constant target portfolio volatility $\sigma_{\text{tgt}}$ (set to 10\% in our empirical evaluation), the realized portfolio weight $w_{t,k}$ allocated to asset $k$ is obtained by scaling the neural network's signal by this leverage factor:
\begin{equation}
    w_{t,k} = \hat{y}_{t,k} \left( \frac{\sigma_{\text{tgt}}}{\sigma_{t,k}} \right) = \hat{y}_{t,k} \cdot \sigma_{\text{tgt}} \cdot \text{vs\_factor}_{t,k}
\end{equation}
Given these target weights, the daily gross strategy return for a specific asset $k$ realized at time $t+1$ is the product of the position taken at the end of day $t$ and the subsequent asset return $r_{t+1,k}$:
\begin{equation}
    R_{t+1, k} = w_{t,k} \cdot r_{t+1,k}
\end{equation}
Assuming an equal risk capital allocation across the $K$ active assets, the aggregate daily gross portfolio return $R_{t+1}^{\text{port}}$ is the cross-sectional average of the individual strategy returns:
\begin{equation}
    R_{t+1}^{\text{port}} = \frac{1}{K} \sum_{k=1}^K R_{t+1, k}
\end{equation}
\paragraph{End-to-end Optimization} Unlike traditional forecasting models that minimize predictive errors, our framework directly optimizes for risk-adjusted economic performance~\cite{enhancing_tsmom}. To train the parameters $\theta$ of the network $f_\theta$, we compute the Sharpe Ratio over a given training sequence of length $T$. Let $\mathbf{R}^{\text{port}} = \{ R_1^{\text{port}}, \dots, R_T^{\text{port}} \}$ represent the sequence of daily portfolio returns. We define the sample estimators for the expected return $\hat{\mathbb{E}}[R]$ and variance $\hat{\text{Var}}[R]$ of the portfolio as:
\begin{subequations}\label{eq:mean_var_port}
\begin{align}
\hat{\mathbb{E}}[R]
&= \frac{1}{T}\sum_{t=1}^{T} R_t^{\text{port}}, \\
\hat{\mathrm{Var}}[R]
&= \frac{1}{T}\sum_{t=1}^{T}\left(R_t^{\text{port}} - \hat{\mathbb{E}}[R]\right)^2.
\end{align}
\end{subequations}
The loss function $\mathcal{L}(\theta)$ is defined as the negative differentiable annualized Sharpe Ratio:
\begin{equation}
\mathcal{L}(\theta)
= -\,\widehat{\mathrm{SR}}[R]
= -\,\frac{\hat{\mathbb{E}}[R]}{\sqrt{\hat{\mathrm{Var}}[R] + \epsilon}}\sqrt{252}.
\label{eq:sr}
\end{equation}
where $252$ represents the approximate number of trading days in a year, and $\epsilon$ is a small constant added for numerical stability. By minimizing this loss, the network explicitly learns representations that maximize expected returns while heavily penalizing variance. Following the regime-robust DeePM framework \cite{deepm_regime_robust} and~\cite{few_shots}, we compute the Sharpe ratio objective on \emph{pooled} portfolio returns concatenating all sequences in the batch, following their argument that this is the best proxy for optimising out-of-sample Sharpe ratio. See the end-to-end optimization pipeline in Figure~\ref{fig:implementation_diagram}.
\paragraph{Net Returns and Breakeven Transaction Costs}
To account for implementation frictions, the net portfolio return $R_{t+1}^{\text{net}}$ is defined by deducting the costs associated with portfolio turnover:
\begin{equation}
    R_{t+1}^{\text{net}} = R_{t+1}^{\text{port}} - \frac{1}{K} \sum_{k=1}^K c_k |w_{t,k} - w_{t-1,k}|
\end{equation}
where $c_k$ represents the proportional transaction cost per unit of traded weight for asset $k$. Because realistic execution costs vary drastically across the asset universe (e.g., highly liquid short-term interest rates versus illiquid agricultural commodities), imposing static ex-ante assumptions for $c_k$ can severely distort cross-sectional performance metrics. Therefore, for model optimization and primary evaluation, we set $c_k = 0$ for all $k$ to assess the pure predictive efficacy of the architectures (gross returns). To evaluate resilience to trading frictions without relying on arbitrary assumptions, we conduct a post-hoc, asset-level breakeven transaction cost analysis. For each asset, we compute the breakeven cost $c_k^*$, which represents the maximum constant friction that specific asset's strategy can endure before its cumulative PnL is driven to zero. The formal mathematical definition of $c_k^*$ and the comprehensive per-asset breakeven results are detailed in Appendix \ref{appdx:asset_level}. As detailed by~\cite{deepm_regime_robust}, we ensemble the positions of the top $10$ seeds based on validation loss.
\subsection{Linear Baselines}

We include a set of linear models as classical and modern baselines to contextualize the performance of deep architectures. The mathematical background can be found in~\ref{appdx:linear}

\paragraph{Autoregressive Model (AR1x).}
The AR1x model \cite{ar1} serves as a minimal temporal benchmark, capturing short-term autocorrelation in returns. This model applies AR(1) per feature, since the input has multiple features. Its performance provides a lower bound on the benefit of incorporating temporal context and highlights whether short-horizon dependence alone is sufficient for profitable trading.

\paragraph{DLinear and NLinear.}
DLinear and NLinear \cite{nlinear} are non-recurrent linear models that apply learned linear mappings to fixed-length input windows. DLinear explicitly decomposes the input into trend and seasonal components, while NLinear operates on normalized inputs. These models have shown strong performance on data with pronounced linear structure or seasonality.

\subsection{Transformer-Based Architectures Without Explicit Recurrence}

Within the Transformer~\ref{appdx:transformer} temporal context, i.e., context window, the model learns relative temporal importance.
However, it does not explicitly encode a temporal state representation and is therefore susceptible to overfitting to outliers.

Transformer-based models are known to perform well in many tasks, but often struggle with financial time-series forecasting~\cite{trans_fail_in_ts1, trans_fail_in_ts2}. In the implementation of the following models, we use parallel offset streams so we obtain dense per-timestep outputs, similar to the output of the other models.

\paragraph{iTransformer.}
The inverted Transformer \cite{itransformer} applies attention across feature dimensions rather than time, treating each feature as a token. While this design improves parameter efficiency, it removes explicit temporal recurrence and relies solely on attention to capture dynamics. 

\paragraph{PatchTST.}
PatchTST \cite{patchtst} segments the input sequence into temporal patches, which are embedded and processed via self-attention. Representing the data in patches inherently smooths the financial time-series and has shown to improve the performance in long-term forecasting \cite{trans_in_ts}. The receptive field is increased to fully exploit the temporal context captured by the patch-based architecture.

\subsection{State-Space and Implicitly Recurrent Models}

\paragraph{Mamba and Mamba2.}
Mamba models \cite{mamba, mamba2} belong to the class of selective state-space models (SSMs)~\ref{appdx:mamba}, which maintain a latent state that is updated recursively over time. Unlike attention-based architectures, SSMs provide an implicit temporal recurrence with linear-time complexity, making them well-suited for long sequences and noisy environments such as financial time series.

At a high level, these models update a hidden state, often based on High-order Polynomial Projection Operators (HiPPO)~\cite{hippo}, which summarizes past information and produces outputs conditioned on the current state. The parameters governing the state evolution are dynamically modulated by neural networks conditioned on the input, enabling adaptive temporal dynamics while preserving computational efficiency.

Mamba2 refines this formulation by simplifying the state transition structure and increasing head dimensionality, leading to improved numerical stability and throughput. In our implementation, we use a static HiPPO-based state transition matrix with a fixed horizon, rather than per-step adaptive horizon jitter. This design choice improves noise tolerance \cite{simple_mamba} and stabilizes learning in the presence of heavy-tailed returns and regime shifts, which are common in financial markets. 

\subsection{Recurrent Models}
\label{sec:recurrent_models}

\paragraph{LSTM.}
Long Short-Term Memory (LSTM)~\ref{appdx:lstm} networks maintain an explicit recurrent state, consisting of a hidden state and a memory cell, that is updated sequentially over time, enabling the model to capture temporal dependencies, but with an exponentially decaying long-horizon temporal state \cite{lstm}. This architecture has proven to be useful in many cases in finance \cite{char_fin_ts1}.

\paragraph{xLSTM.}

xLSTM \cite{xlstm} extends the classical LSTM by introducing exponential gating and stabilized memory normalization to improve long-range information retention and gradient flow~\ref{appdx:lstm}. While standard LSTMs rely on sigmoid gates that may saturate and induce premature forgetting, xLSTM replaces these with exponentiated gate activations followed by normalization, yielding approximately linear behavior over a wider dynamic range. This modification mitigates vanishing memory effects and allows the model to retain rare but economically meaningful signals.

xLSTM comprises two variants: scalar LSTM (sLSTM), which maintains a scalar memory state updated via normalized exponential gates, and matrix LSTM (mLSTM), which generalizes the memory state to a matrix-valued representation, enabling higher memory capacity and associative recall through key–value storage mechanisms. This design increases representational richness and improves scalability compared to classical recurrent architectures.

From a financial perspective, the ability to preserve temporally distant but informative signals and adaptively revise memory states is particularly relevant in low signal-to-noise and regime-dependent environments \cite{lstm_on_fin_ts}.

\paragraph{Patch sLSTM (PsLSTM).}
Patch sLSTM \cite{pslstm} integrates the patching strategy of PatchTST with the recurrent inductive bias of sLSTM. Given a multivariate time series of length $L$ with $d$ channels, each channel is treated as an independent univariate sequence and segmented into non-overlapping temporal patches:
\begin{equation}
    \tilde{x}_p^{(i)} = \text{Patch}\left(x_{t:t+\ell-1}^{(i)}\right), \quad (i = 1,\dots,d)
\end{equation}
where $\ell$ denotes the patch length.

Each patch embedding $\tilde{x}_p^{(i)}$ is then processed by an sLSTM, with \emph{shared parameters across channels}:
\begin{equation}
    h_p^{(i)} = \text{sLSTM}(h_{p-1}^{(i)}, \tilde{x}_p^{(i)}).
\end{equation}
This design preserves channel independence while enforcing parameter sharing, preventing premature feature mixing and improving generalization.

By operating at the patch level, PsLSTM reduces sensitivity to high-frequency noise and allows the recurrent mechanism to focus on medium-term temporal structure. The exponential gating of sLSTM further enhances memory persistence across patches, enabling the model to capture regime-level dynamics and rare events. After recurrent processing, hidden states across channels are flattened and projected to form the final prediction.

Patch sLSTM thus combines the noise robustness and efficiency of patch-based modeling with the long-range memory advantages of exponential-gated recurrence, which may be advantageous in financial time series characterized by non-stationary and bad signal-to-noise ratio.

\subsection{Hybrids}

Several hybrid architectures are considered to improve robustness in financial time series by enhancing the signal-to-noise ratio and stabilizing temporal state updates~\ref{appdx:hybrid}.

Variable Selection Networks (VSNs)~\ref{appdx:vsn}, inspired by the Temporal Fusion Transformer, are used to perform feature-wise nonlinear embedding and dynamic soft selection of relevant covariates at each time step. This mechanism adaptively suppresses noisy or uninformative features.

Another strategy to improve robustness is the inclusion of an LSTM-based temporal encoder prior to the main model. By explicitly maintaining a recurrent state, this preprocessing stage filters high-frequency noise and stabilizes downstream representations.

\paragraph{VSN+LSTM (VLSTM).}
VLSTM combines a VSN with an LSTM encoder to construct a compact temporal state representation. This was the core component of the X-Trend architecture used for constructing sequence representations for few-shot learning in financial time series~\cite{few_shots}.
The VSN produces dynamically weighted feature embeddings, which are then processed sequentially by an LSTM to aggregate long-range temporal information. 

\paragraph{VSN--Mamba2.}
This hybrid augments Mamba2 with a VSN to separate feature selection from temporal modeling. The VSN filters noisy covariates before passing the selected representation to the recurrent state-space model, improving robustness in noisy financial environments.

\paragraph{LSTM + PatchTST (LPatchTST).}
This architecture combines explicit recurrence with attention by using an \textit{LSTM as a channel-wise temporal denoiser} prior to PatchTST. The LSTM stabilizes per-channel representations, while PatchTST aggregates medium- and long-range dependencies across denoised temporal patches.

\paragraph{VSN + xLSTM (VxLSTM).}
In this hybrid, VSN-selected representations are directly fed into an xLSTM. The matrix-valued memory of xLSTM enables the model to capture higher-order temporal interactions and long-range dependencies beyond the capacity of vector-based recurrent architectures.

\subsection{Complete Structured Model}

\paragraph{Temporal Fusion Transformer (TFT).}
TFT \cite{tft} integrates gated recurrent layers with interpretable attention mechanisms. A recurrent encoder captures local temporal dynamics, where LSTM is used as the recurrent encoder, while multi-head attention aggregates information across time:
\begin{equation}
    \hat{y}_t = \text{Attn}(\text{LSTM}(x_{1:t})).
\end{equation}
Variable selection networks, static covariate encoders, and gating mechanisms further improve robustness, making TFT a strong benchmark for time-series forecasting and specifically financial forecasting \cite{mom_trans}.

%% file: Sections/benchmark.tex
This section presents a comprehensive evaluation of the out-of-sample performance of the considered models across multiple market regimes, performance metrics, and computational dimensions, Appendix~\ref{appdx:measures}. We focus on (i) risk-adjusted returns across subperiods, (ii) aggregate return performance and statistical significance, (iii) downside and tail-risk characteristics, (iv) robustness to seed selection and experimental budget, and (v) the trade-off between predictive performance and computational complexity.

\subsection{Data Description}
\label{sec:data_description}

Our empirical analysis was conducted on a diversified cross-asset futures and currency dataset between 1990 and 2024~\cite{pinnacle_data}. We used a walk-forward backtesting approach: our initial training period was 1990–2010, followed by an out-of-sample testing period from 2010–2024 evaluated using a 5-year rolling window. The futures data comprises of instruments from five asset classes: bonds, commodities, energy, foreign exchange, and equity indices. Daily closing prices were used to construct returns and predictive features. For futures, we use continuous contracts formed by linking adjacent maturities using a ratio-adjusted backwards methodology (i.e., back-adjusted to remove roll-induced price jumps).
A detailed description of data construction and exploratory analysis is provided in Appendix~\ref{appdx:eda}.

The dataset~\cite{pinnacle_data} exhibits several well-documented stylized facts of financial time series, including heavy-tailed return distributions, volatility clustering, and strong deviations from Gaussianity. These properties are illustrated formally in Appendix~\ref{appdx:eda}, where we reported distributional diagnostics (e.g., QQ-plots) and volatility dynamics. As such, the dataset provides a realistic and challenging benchmark for evaluating nonlinear forecasting architectures in a cross-asset setting.

Although the empirical evidence is robust within our cross-asset benchmark, the results should be interpreted as conditional on the specific dataset and period considered. Extending the analysis to alternative markets and sampling frequencies would further clarify the external validity of the findings.


\subsection{Performance Across Market Regimes}

\begin{table*}[t]
\centering
\caption{Out-of-sample Sharpe Ratio by subperiod. Annual Sharpe Ratio in Table~\ref{tab:performance_by_year_extended}}
\resizebox{\textwidth}{!}{
\begin{tabular}{lrrrrr}
\toprule
Strategy & 2010-2024 & 2015-2024 & 2010-2014 & 2015-2019 & 2020-2024\\
\midrule
AR1x        & 0.77 & 0.71 & 0.89 & 0.06 & 1.35\\
AR$n$x      & 0.63 & 0.55 & 0.80 & -0.01 & 1.11\\
DLinear     & 0.64 & 0.64 & 0.65 & 0.00 & 1.28\\
NLinear     & 0.52 & 0.48 & 0.61 & 0.20 & 0.77 \\
Mamba       & 0.36 & 0.28 & 0.53 & -0.01 & 0.56 \\
Mamba2      & 0.80 & 0.87 & 0.68 & 0.19 & 1.54\\
VSN+Mamba2  & 1.10 & 1.14 & 1.01 & 0.54 & 1.74\\
LSTM        & 1.48 & 1.33 & 1.78 & 1.60 & 1.07\\
VLSTM       & \textbf{2.40} & \textbf{2.24} & \textbf{2.70} & \textbf{2.61} & 1.88\\
VxLSTM      & 1.69 & 1.61 & 1.85 & 1.48 & 1.74\\
xLSTM       & 1.79 & 1.84 & 1.68 & 1.69 & 2.00\\
PsLSTM      & 1.78 & 1.78 & 1.77 & 1.81 & 1.76\\
PatchTST    & 0.75 & 0.79 & 0.76 & 0.51 & 1.07\\
LPatchTST   & \underline{2.29} & \underline{2.18} & 2.37 & 2.05 & \textbf{2.30}\\
TFT         & 2.27 & 2.08 & \underline{2.66} & \underline{2.08} & \underline{2.08} \\
iTransformer& 0.38 & 0.28 & 0.58 & 0.07 & 0.50\\
\bottomrule
\end{tabular}
}
\label{tab:performance_by_period_extended}
\end{table*}

Table~\ref{tab:performance_by_period_extended} reports out-of-sample Sharpe ratios~\cite{sharpe_ratio} aggregated over multiple overlapping horizons from 2010 to 2024, while Table~\ref{tab:performance_by_year_extended} presents annual Sharpe ratios. Taken together, these results enable an evaluation of year-to-year variability and medium-horizon robustness across distinct market regimes, including the post-GFC recovery, the low-volatility expansion of the mid-2010s, and the elevated-uncertainty environment following 2020.

Several systematic patterns emerge.

First, deep nonlinear sequence models substantially outperform linear benchmarks on most aggregated horizons. While linear specifications such as AR1x, AR$n$x, DLinear, and NLinear occasionally achieve strong single-year Sharpe ratios—particularly during high-volatility years such as 2020—their performance is highly variable across time. Their long-horizon averages over 2010–2025 remain materially below those of the strongest nonlinear architectures. This instability is consistent with the limited representational flexibility of linear dynamics in environments characterized by non-stationarity, regime shifts, and low signal-to-noise ratios.

In contrast, gated recurrent and hybrid sequence models exhibit both higher average Sharpe ratios and greater intertemporal consistency. The LSTM already delivers strong performance (1.48 over 2010–2025), but its variance across years remains non-negligible. Enhanced recurrent architectures improve further. In particular, VLSTM achieves a 2010–2025 Sharpe ratio of 2.40 and maintains strong performance across subperiods, including 2.24 over 2015–2025 and 1.88 over 2020–2025. Similarly, LPatchTST achieves 2.29 over 2010–2025 and remains stable across all aggregated windows, including 2.30 in the post-2020 regime. The Temporal Fusion Transformer (TFT) also demonstrates robust performance, with 2.27 over 2010–2025 and consistent strength across medium-horizon splits.

These results suggest that architectures combining adaptive gating, representation compression, and structured temporal abstraction are better suited to financial data than either purely linear models or attention-only baselines. The year-by-year breakdown further reveals that top-performing models rarely collapse entirely in adverse years; rather, performance degrades moderately while remaining economically meaningful. This robustness is particularly visible during volatile periods such as 2020–2022.

State-space models such as Mamba and Mamba2 display more heterogeneous behavior. While certain years exhibit strong Sharpe ratios (notably 2020 and 2022 for Mamba2), their aggregated performance remains moderate (0.80 and 0.36 over 2010–2025 for Mamba2 and Mamba, respectively). Augmenting Mamba2 with a Variable Selection Network improves medium-horizon averages (1.10 over 2010–2025), indicating that explicit feature conditioning partially mitigates instability, though it does not close the gap to the strongest recurrent or hybrid models.

Transformer-based patching approaches show mixed results. PatchTST achieves moderate long-run averages (0.75 over 2010–2025), but exhibits higher sensitivity to specific years. In contrast, LPatchTST, which augments patching with stronger sequence modeling, delivers consistently superior and more stable results, suggesting that patch segmentation alone is insufficient without robust temporal state encoding.

Finally, xLSTM-based architectures demonstrate a compelling balance between performance and stability. The xLSTM achieves a Sharpe ratio of 1.79 over 2010–2025, improving to 2.00 in the 2020–2025 period. VxLSTM yields comparable results (1.69 over 2010–2025), while PsLSTM achieves 1.78. Importantly, these models maintain Sharpe ratios near or above 1.5 across most aggregated horizons, indicating resilience to changing volatility regimes. Their year-level profiles show fewer extreme drawdowns relative to classical LSTM, consistent with the hypothesis that enriched state representations and alternative gating mechanisms enhance adaptability in non-stationary environments.

Taken together, the annual and aggregated results reinforce the central hypothesis of the paper: successful financial forecasting architectures benefit from adaptive memory mechanisms, representation compression, and temporally stable state evolution. Models that incorporate structured gating and persistent state representations dominate both linear baselines and generic state-space formulations across nearly all evaluation horizons. As emphasized throughout, these conclusions remain conditional on the dataset and evaluation framework considered; nevertheless, the consistency across multiple temporal aggregations provides evidence that the observed performance differentials are not driven solely by isolated years or singular market events.

\subsection{Aggregate Return Performance and Statistical Significance}
\label{sub:agg_performanc}

Table~\ref{tab:performance_main} reports full-sample out-of-sample performance under a volatility-targeting constraint of 10\%, presenting compound annual growth rates (CAGR)~\cite{elton_mpt}, annualized returns (Ann. Ret.), heteroskedasticity and autocorrelation consistent $t$-statistics
($t$ HAC)~\cite{newey_west}, hit rate (Hit)~\cite{pesaran_timmermann}, turnover, turnover as a multiple of gross market value (xGMV)~\cite{vol_target}, and additional diagnostics relative to a passive long-only benchmark: the information ratio (Info. Ratio), HAC $t$-statistic relative to passive ($t$ HAC v Passive), and correlation with passive returns (Corr. v Passive). Collectively, these measures capture economic magnitude, statistical reliability, trading intensity, and incremental value relative to buy-and-hold exposure, see Appendix~\ref{appdx:metrics}.

VLSTM delivers the strongest overall performance within this framework. It achieves a 23.9\% annualized return with a Sharpe ratio of 2.40 (table~\ref{tab:performance_by_period_extended}), exceeding both linear and alternative deep learning benchmarks. Its HAC-adjusted $t$-statistic of 8.81 indicates high statistical reliability under heteroskedasticity- and autocorrelation-consistent inference. The hit rate of 58.8\% suggests persistent directional accuracy. Relative to the passive benchmark, VLSTM attains an information ratio of 0.854 and an HAC $t$-statistic of 3.31, indicating statistically distinguishable excess performance. Its correlation with passive returns (0.404) implies partial independence from broad market exposure and meaningful diversification potential.

The hybrid LPatchTST model achieves comparable economic performance, with a Sharpe ratio of 2.29 and a CAGR of 25.5\%. Passive-relative metrics remain elevated (information ratio 0.702; $t$ HAC 2.74), though modestly below VLSTM. Similarly, the Temporal Fusion Transformer (TFT) delivers strong absolute and relative performance (Sharpe 2.27), reinforcing the importance of structured sequence representations and adaptive gating mechanisms. Across these leading architectures, elevated Sharpe ratios coincide with statistically significant passive-relative improvements, suggesting that gains are not merely attributable to implicit market timing or leverage effects under volatility targeting.

LSTM-based variants, including PsLSTM, VxLSTM, and xLSTM, substantially outperform linear baselines. Notably, xLSTM achieves a Sharpe ratio of 1.79 with a comparatively moderate turnover (482), resulting in one of the strongest passive-relative diagnostics (information ratio 0.798; $t$ HAC 2.90). This combination of competitive returns and reduced trading intensity suggests improved efficiency in signal extraction relative to classical LSTM, which requires nearly double the turnover to achieve similar Sharpe ratios. These findings are consistent with the view that enriched state representations can improve the signal-to-trade ratio in noisy financial environments.

Linear benchmarks (AR1x, AR$n$x, DLinear) exhibit Sharpe ratios below one and comparatively small HAC $t$-statistics. Passive-relative metrics are near zero or negative, and correlations with the buy-and-hold benchmark remain moderate. While these models occasionally benefit from favorable return persistence, their aggregate performance indicates limited capacity to extract stable predictive structure under non-stationarity.

State-space models display heterogeneous behavior. Mamba2 reduces trading intensity relative to most deep sequence models (turnover 233) but achieves only moderate economic performance (Sharpe 0.80). Augmenting with a Variable Selection Network improves both Sharpe ratio (1.10) and passive-relative metrics, indicating that explicit feature conditioning enhances stability. Nevertheless, their aggregate performance remains below that of recurrent and hybrid architectures.

The inclusion of iTransformer provides an informative contrast. It exhibits by far the lowest turnover (36) and xGMV (9.2), indicating a highly conservative trading profile with minimal portfolio rebalancing. However, this low implementation intensity coincides with weak economic performance (Sharpe 0.38) and statistically insignificant passive-relative diagnostics. This pattern suggests that extreme turnover reduction may reflect under-reactivity to evolving return signals in non-stationary markets. In the present setting, reduced trading alone does not generate economic value; rather, successful models appear to balance adaptive responsiveness with controlled trading intensity.

\begin{figure*}[t]
\centering
\includegraphics[width=0.8\textwidth]{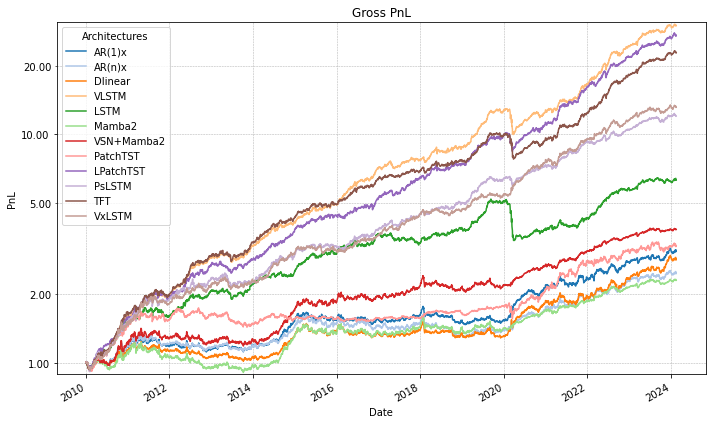}
\caption{Performance comparison across models 10\% volatility-rescaled gross PnL. 
}
\label{fig:performance_comparison}
\end{figure*}

Figure~\ref{fig:performance_comparison} visually corroborates these quantitative findings. Leading sequence-based models dominate cumulative PnL trajectories while maintaining relatively smooth performance paths, indicating that elevated Sharpe ratios reflect persistent incremental returns rather than isolated return episodes. As throughout, these findings remain conditional on the dataset and backtesting design considered.

\begin{figure*}[t]
\centering
\includegraphics[width=0.7\textwidth]{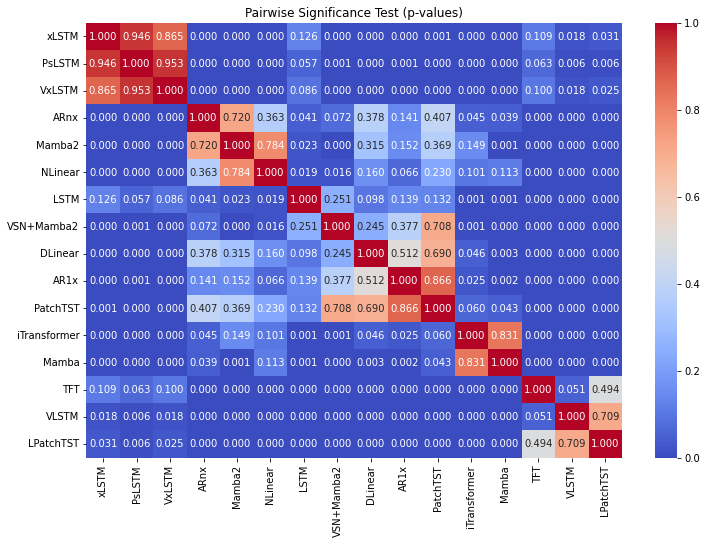}
\caption{Pairwise Significance Test based on p-value
}
\label{fig:pairwise_significance_pval}
\end{figure*}

\begin{figure*}[t]
\centering
\includegraphics[width=0.7\textwidth]{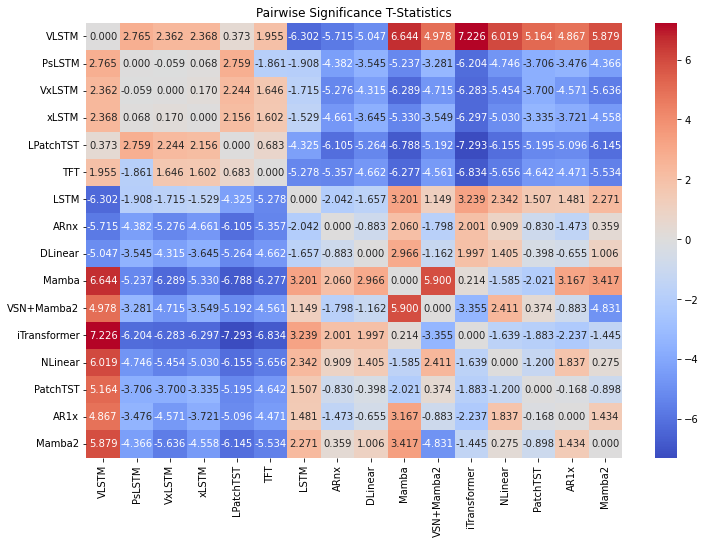}
\caption{Pairwise Significance Test based on t-statistics
}
\label{fig:pairwise_significance_tstat}
\end{figure*}

To assess whether the observed differences in risk-adjusted performance are statistically distinguishable across architectures, we conduct pairwise paired $t$-tests on the daily portfolio return series. For every pair of models, we compute both the $t$-statistic and the corresponding two-sided $p$-value using the standard paired $t$-test (scipy.stats.ttest\_rel). The resulting matrices are shown in Figure~\ref{fig:pairwise_significance_pval} and~\ref{fig:pairwise_significance_tstat}.

The $p$-value heatmap reveals several clear clusters of statistical indistinguishability. The three xLSTM-family models (xLSTM, PsLSTM and VxLSTM) are mutually non-significant ($p\in\{0.865,0.946,0.953\}$), consistent with their shared architectural lineage. A second, weaker cluster appears among the linear and state-space baselines (ARnx, Mamba2, NLinear), with pairwise $p$-values ranging from 0.36 to 0.78. PatchTST likewise shows elevated $p$-values against several of these lower-performing models. At the opposite end of the spectrum, Mamba and iTransformer are statistically indistinguishable from each other ($p=0.831$), while LPatchTST exhibits relatively high $p$-values against both VLSTM and TFT.

The complementary $t$-statistic heatmap quantifies the magnitude and direction of these differences. The six strongest architectures—VLSTM, LPatchTST, TFT, xLSTM, PsLSTM and VxLSTM—produce large positive $t$-statistics (typically $|t|>4$–$7$) against virtually the entire set of weaker models (LSTM, the linear baselines, Mamba-family variants, PatchTST, iTransformer, etc.). Conversely, the same weaker models produce large negative $t$-statistics when compared with the top tier. Within the leading group the $t$-statistics are modest, confirming that the economic ranking among the best models is stable but that several pairwise differences do not reach conventional significance thresholds.

Taken together, the pairwise tests reinforce the hierarchical ordering already visible in the aggregate metrics: the top-performing architectures are reliably distinguishable from the lower tier, while differences inside each tier are often statistically insignificant. This provides formal support for the claim that the observed performance gaps are not artefacts of sampling variation.


\begin{table*}[t]
\centering
\scriptsize
\setlength{\tabcolsep}{4pt}
\caption{2010--2025 Gross return performance, statistical significance, and passive-relative diagnostics (volatility-targeted at 10\%).} 
\begin{tabular}{lccccccccc}
\toprule
Model &
CAGR &
Ann. Ret. &
$t$ (HAC) &
Hit &
Turnover &
xGMV & 
Info. Ratio & 
$t$ (HAC) v Passive & 
Corr. v Passive \\
\midrule
Passive            & 0.0435 & 0.048 & 1.65 & 0.531 & -- & -- & -- & -- & -- \\
AR1x               & 0.0813 & 0.077 & 3.12 & 0.537 & 353.64 & 90.421  & -0.0086 & -0.0305 & 0.3533 \\
AR$n$x            & 0.0647 & 0.063 & 2.52 & 0.538 & 280.66 & 69.525 & -0.0830 & -0.3011 & 0.4325 \\
DLinear            & 0.0750 & 0.064 & 2.87 & 0.539 & 278.41 & 75.282  & 0.0141 & 0.0501 & 0.2612 \\
NLinear            & 0.0541 & 0.052 & 2.10 & 0.543 & 178.87 & 46.786  & -0.0812 & -0.2896 & 0.4806 \\
Mamba            & 0.0266 & 0.036 & 1.15 & 0.523 & \underline{128.99} & \underline{34.673}  & -0.2439 & -0.8925 & \underline{0.2289} \\
Mamba2             & 0.0587 & 0.080 & 2.31 & 0.546 & 233.00 & 58.164 & -0.0901 & -0.3246 & \textbf{0.2220} \\
VSN+Mamba2         & 0.0967 & 0.110 & 3.65 & 0.555 & 329.11 & 78.842 & 0.1091 & 0.3936 & 0.2821 \\
LSTM               & 0.1351 & 0.148 & 4.56 & 0.554 & 948.08 & 225.769 & -0.0637 & -0.2303 & 0.2816 \\
VLSTM             & \textbf{0.2632} & \textbf{0.240} & \underline{8.81} & \textbf{0.588} & 966.86 & 218.369 & \textbf{0.8539} & \textbf{3.3071} & 0.4042 \\
VxLSTM          & 0.1937 & 0.169 & 6.89 & 0.574 & 775.88 & 159.438 & 0.4666 & 1.6727 & 0.5069 \\
xLSTM              & 0.1907 & 0.179 & 6.64 & 0.568 & 482.62 & 91.924  & \underline{0.7984} & \underline{2.9042} & 0.6274 \\
PsLSTM             & 0.1923 & 0.178 & 7.17 & 0.560 & 884.61 & 198.167 & 0.3886 & 1.5055 & 0.4695 \\
PatchTST           & 0.0855 & 0.075 & 3.31 & 0.538 & 613.85 & 195.318 & -0.199 & -0.7188 & 0.5519 \\
LPatchTST    & \underline{0.2548} & \underline{0.229} & \textbf{8.96} & 0.572 & 949.09 & 210.144 & 0.7016 & 2.7387 & 0.3187 \\
TFT                & 0.2398 & 0.227 & 8.13 & \underline{0.584} & 912.81 & 223.231 & 0.6665 & 2.5487 & 0.3888 \\
iTransformer       & 0.0308 & 0.038 & 1.26 & 0.529 & \textbf{36.32} & \textbf{9.203}  & -0.1539 & -0.5563 & 0.4855 \\
\bottomrule
\end{tabular}
\label{tab:performance_main}
\end{table*}





\subsection{Downside Risk and Tail Behavior}
\label{sub:downside}

While strong average performance is economically relevant, robustness to adverse market conditions is particularly important in financial applications. Table~\ref{tab:risk_metrics} reported downside and tail-risk metrics, including maximum drawdown (Max DD), Calmar ratio (Calmar), worst three-month Sharpe ratio (Worst 3m Sharpe), minimum annual Sharpe ratio (Min Ann. Sharpe), and 5\% conditional value at risk (CVaR 5\%).

VLSTM and LPatchTST exhibited comparatively moderate drawdowns alongside relatively high Calmar ratios (1.15 and 1.64, respectively). VLSTM achieved a worst three-month Sharpe ratio of -3.68 and the second lowest CVaR among the evaluated models (0.0137) after LPatchTST (0.0136). These results were consistent with comparatively milder tail losses within the sample period.

The smallest maximum drawdown was observed for VxLSTM (-11.8\%), accompanied by the highest Calmar ratio (1.64). However, this configuration generated lower average returns relative to VLSTM in the aggregate performance analysis, suggesting a more conservative return profile within the evaluation framework.

In contrast, standard LSTM and PatchTST architectures experienced larger drawdowns and weaker worst-period performance. This pattern indicated greater sensitivity to extreme market movements during the sample period.

Overall, VLSTM combined comparatively strong average performance with moderate downside risk measures. However, LPatchTST and xLSTM were the most robust and were able to keep a favorable tail behavior. Within the dataset considered, this balance suggested a favorable trade-off between return generation and tail-risk exposure. As throughout, these findings should be interpreted as conditional on the sample period and backtesting design.







\begin{table*}[t]
\centering
\caption{Downside risk and tail behavior (gross returns).}
\resizebox{\textwidth}{!}{
\begin{tabular}{lccccc}
\toprule
Model &
Max DD &
Calmar &
Worst 3m Sharpe &
Min Ann. Sharpe &
CVaR 5\% \\
\midrule
Passive            & -0.308 & 0.14 & -6.11 & -1.53 & 0.0144 \\
AR1x               & -0.167 & 0.49 & -3.92 & -0.59 & 0.0147 \\
AR$n$x             & -0.206 & 0.31 & -4.52 & -0.90 & 0.0148 \\
DLinear            & -0.180 & 0.42 & -4.93 & -0.94 & 0.0149 \\
NLinear            & -0.227 & 0.24 & -4.54 & -1.19 & 0.0151 \\
Mamba              & -0.247 & 0.11 & -4.76 & -0.54 & 0.0152 \\
Mamba2             & -0.263 & 0.22 & -4.06 & -0.49 & 0.0149 \\
VSN+Mamba2         & -0.163 & 0.59 & -4.00 & -0.63 & 0.0148 \\
LSTM               & -0.342 & 0.40 & -5.15 & -1.51 & 0.0143 \\
VLSTM              & -0.229 & 1.15 & -3.68 & \underline{-0.10} & \underline{0.0137} \\
VxLSTM          & \textbf{-0.118} & \textbf{1.64} & -3.70 & -1.31 & 0.0139 \\
xLSTM              & -0.141 & 1.35 & \textbf{-3.57} & -0.28 & 0.0141 \\
PsLSTM             & \underline{-0.127} & \underline{1.51} & \underline{-3.60} & -0.89 & 0.0142 \\
PatchTST           & -0.184 & 0.47 & -5.33 & -1.19 & 0.0151 \\
LPatchTST      & -0.155 & \textbf{1.64} & -3.64 &  \textbf{0.59} & \textbf{0.0136} \\
TFT                & -0.232 & 1.03 & -3.87 & -0.14 & 0.0141 \\
iTransformer       & -0.259 & 0.12 & -3.93 & -1.16 & 0.0154 \\
\bottomrule
\end{tabular}}
\label{tab:risk_metrics}
\end{table*}


\subsection{Breakeven Transaction Cost}


Breakeven transaction cost analysis, Appendix~\ref{appdx:asset_level}, reveals substantial cross-asset heterogeneity. The tables report annualised, volatility-rescaled gross and net returns together with annualised turnover and the implied breakeven transaction cost $c^*$ in basis points (bps). Note that the transaction costs of the futures we are considering are typically 1 bps and sometimes less.

For VLSTM (Table~\ref{tbl:vlstm_tcost}), several agricultural contracts (e.g., Lumber, Oats elec, and Milk III) exhibit high gross returns but also relatively large breakeven costs (exceeding 20 bps). These contracts are comparatively illiquid therefore they are expected to have high transaction costs and scalability is limited despite their strong gross profitability.
A broad middle group demonstrates moderate profitability, with breakeven costs in the range of 5–10 bps.
At the lower end, high-turnover contracts (e.g., US 2Y Note Composite Bond and Euro Schatz Bond) display very small breakeven costs, indicating that profitability is quickly eroded by transaction costs. These are amongst the most liquid contracts; therefore, it is expected that they have tight spreads. However, they can also be traded in high volumes.
Finally, a small subset of contracts generates negative gross returns, resulting in negative $c^*$.

For xLSTM (Table~\ref{tbl:xlstm_tcost}), the strongest contracts again include Lumber, Oats elec, Milk III, with even higher breakeven costs for some assets (e.g., Lumber at 33.9 bps), once again, noting that these are highly illiquid contracts. Notably, xLSTM achieves materially lower turnover for several equity and bond contracts (e.g., ES, ZN), leading to higher breakeven margins despite similar gross returns. However, the lower tail contains more negative gross performers relative to VLSTM.

Overall, xLSTM appears more transaction-cost efficient in several liquid contracts due to reduced turnover, while VLSTM delivers broader cross-sectional profitability.

\begin{figure*}[p]
    \centering

    \begin{subfigure}[b]{0.49\linewidth}
        \includegraphics[width=\linewidth]{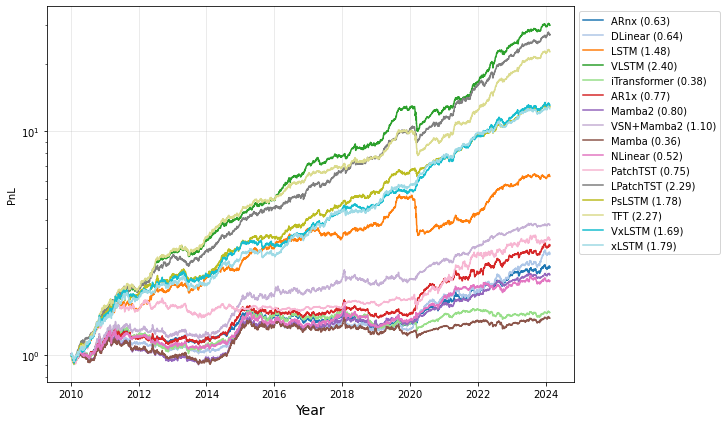}
        \caption{0 bps}
    \end{subfigure}
    \hfill
    \begin{subfigure}[b]{0.49\linewidth}
        \includegraphics[width=\linewidth]{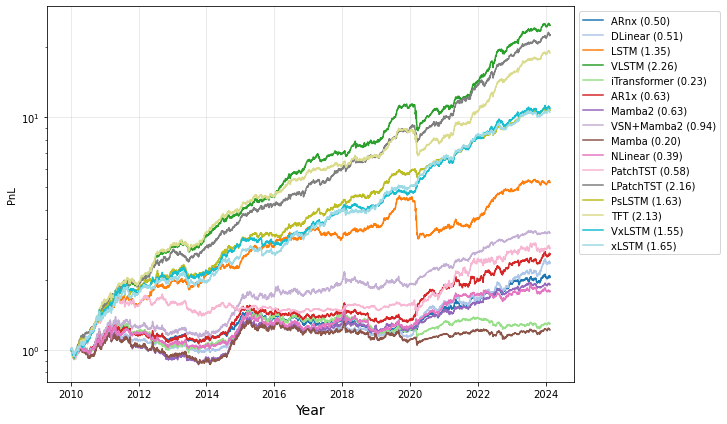}
        \caption{0.5 bps}
    \end{subfigure}

    \vspace{0.3cm}

    \begin{subfigure}[b]{0.49\linewidth}
        \includegraphics[width=\linewidth]{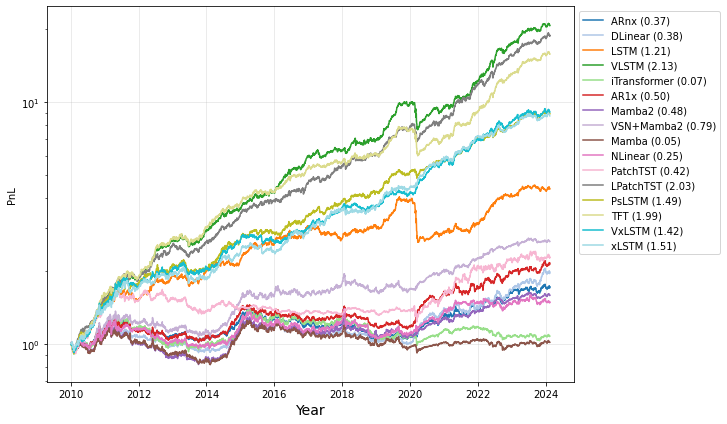}
        \caption{1 bps}
    \end{subfigure}
    \hfill
    \begin{subfigure}[b]{0.49\linewidth}
        \includegraphics[width=\linewidth]{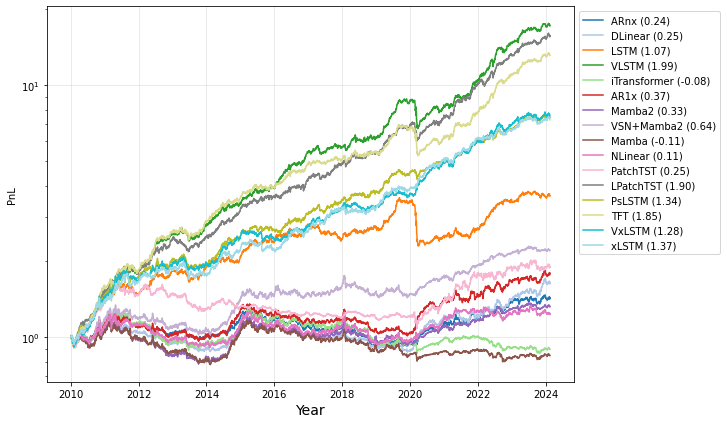}
        \caption{1.5 bps}
    \end{subfigure}

    \vspace{0.3cm}

    \begin{subfigure}[b]{0.49\linewidth}
        \includegraphics[width=\linewidth]{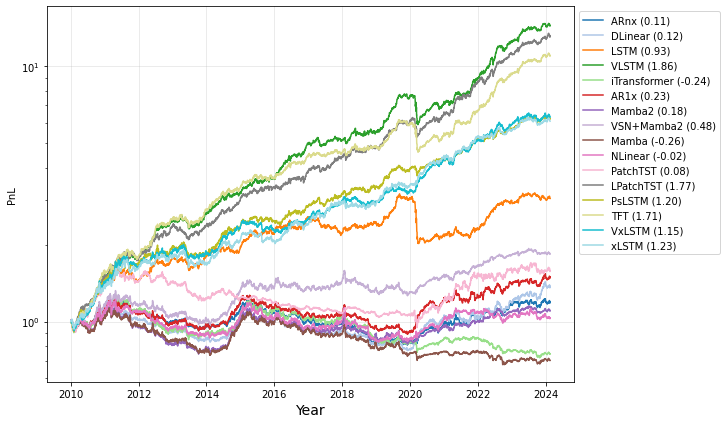}
        \caption{2 bps}
    \end{subfigure}
    \hfill
    \begin{subfigure}[b]{0.49\linewidth}
        \includegraphics[width=\linewidth]{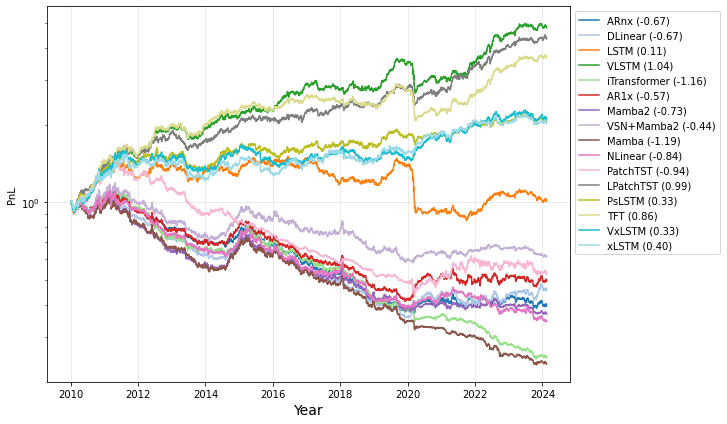}
        \caption{5 bps}
    \end{subfigure}

    \caption{PnL for different transaction costs. Sharpe ratio in brackets.}
    \label{fig:net_pnl_tc}
\end{figure*}

To complement the asset-level breakeven analysis of Appendix~\ref{appdx:asset_level} and to address portfolio-level implementability under realistic trading frictions, we recompute the full out-of-sample equity curves after deducting a constant proportional transaction cost $c$ (in basis points) from every trade. Because the strategies trade liquid commodity, equity-index, bond and FX futures, we examine the economically relevant range $c\in\{0,0.5,1,1.5,2,5\}$\,bps. Figure~\ref{fig:net_pnl_tc} shows the resulting 10\%-volatility-rescaled cumulative PnL for every architecture.

Across the entire range the ranking of the leading models remains essentially unchanged. VLSTM continues to deliver the highest terminal wealth, followed closely by LPatchTST and then by TFT. After a modest gap the three gated recurrent variants VxLSTM, xLSTM and PsLSTM form a second tier whose curves lie almost on top of one another. A further clear separation places the vanilla LSTM next, after which the remaining architectures (linear baselines, PatchTST, Mamba2 and its VSN-augmented counterpart, etc.) cluster together. At the bottom of the hierarchy sit Mamba2 and iTransformer, both of which lose virtually all of their gross-PnL advantage once even modest costs are introduced.

Importantly, the relative ordering is stable already at the lowest non-zero cost levels that are typical of electronic futures markets. Consequently the superiority of the best-performing architectures is not an artefact of a zero-cost assumption; the same models that dominate on a gross-Sharpe basis also remain the most profitable after realistic transaction costs are deducted.

\subsection{Robustness to Seed Selection and Experimental Budget}

Table~\ref{tab:small_seed} evaluated the robustness of the main findings to a substantially reduced experimental budget. Whereas the primary results in Sections~\ref{sub:agg_performanc} and~\ref{sub:downside} were obtained by averaging performance over the top 10 seeds, based on validation loss, from 50 independent runs, this table reported results using only 25 random seeds and averaging over the top 5 realizations. This configuration represented a noisier and more resource-constrained evaluation regime, intended to assess whether model rankings and economic patterns persisted under weaker aggregation.

Across models, performance levels and relative ordering remained broadly stable. VLSTM again achieved the highest absolute and risk-adjusted returns, with a Sharpe ratio of 2.40 and an HAC-adjusted $t$-statistic of 8.86, closely aligned with its full-budget estimates. The TFT and VxLSTM also maintained Sharpe ratios above 1.9 with statistically distinguishable HAC $t$-statistics, while the linear AR1x benchmark continued to exhibit comparatively weak performance across return metrics. This stability in relative rankings was consistent with the interpretation that performance differences were not solely driven by extensive seed exploration or aggressive averaging.

Downside and tail-risk characteristics exhibited similar patterns. Maximum drawdowns, Calmar ratios, and worst three-month returns remained quantitatively comparable to those observed under the full-budget evaluation. VLSTM and VxLSTM maintained elevated Calmar ratios (1.66 and 1.88, respectively), indicating a similar balance between return generation and drawdown magnitude as in the primary specification. The 5\% Conditional Value-at-Risk (CVaR) did not materially deteriorate under reduced seed averaging, suggesting that tail-risk characteristics were not highly sensitive to seed selection within the tested range.

Overall, the reduced-seed experiment suggested that both economic performance and downside risk measures were reasonably stable to substantial reductions in experimental budget. The persistence of model rankings and risk profiles under noisier evaluation conditions was consistent with the view that the observed performance differences reflected structural properties of the architectures within this benchmark, rather than purely favorable random initialization or extreme-seed selection. As throughout, these conclusions remain conditional on the specific dataset and backtesting protocol employed.

\begin{table*}[t]
\small
\setlength{\tabcolsep}{4pt} 
\centering
\caption{Reduced-seed benchmark (25 runs, top 5 seeds selected).}
\scriptsize
\begin{tabular}{lccccc|lccccc}
\toprule
\multicolumn{6}{c|}{\textbf{Return Performance}} & \multicolumn{6}{c}{\textbf{Risk and Downside Metrics}} \\
\midrule
Model & CAGR & Ann. Ret. & Sharpe & $t$ (HAC) & Hit &
Model & Max DD & Calmar & Worst 3m & Min Ann. & CVaR 5\% \\
\midrule
AR1x & 0.084 & 0.085 & 0.854 & 3.208 & 0.537 &
AR1x & -0.165 & 0.506 & -4.068 & -0.524 & 0.0147 \\
VLSTM & 0.264 & 0.240 & 2.397 & 8.857 & 0.589 &
VLSTM & -0.159 & 1.664 & -3.730 & -0.250 & 0.0139 \\
TFT & 0.251 & 0.229 & 2.290 & 8.575 & 0.586 &
TFT & -0.210 & 1.196 & -3.499 & -0.175 & 0.0141 \\
VxLSTM & 0.203 & 0.190 & 1.898 & 7.236 & 0.578 &
VxLSTM & -0.108 & 1.878 & -3.718 & -1.064 & 0.0138 \\
\bottomrule
\end{tabular}
\label{tab:small_seed}
\end{table*}



\subsection{Discussion}
\label{sec:discussion}

The empirical results presented across return, risk, tail, cost, and robustness diagnostics suggest several consistent patterns regarding the inductive biases required for successful financial time-series modeling. As emphasized throughout, these interpretations remain conditional on the dataset and evaluation framework considered, as discussed in Section~\ref{sec:data_description}.

The dataset exhibits non-stationarity, heavy tails, low signal-to-noise ratios, and pronounced deviations from Gaussianity -- stylized facts widely documented in financial return series. To the extent that other financial time series share these characteristics, similar architectural considerations may apply.

First, architectures maintaining explicit recurrent state representations consistently outperformed purely attention-based models across most performance dimensions. This advantage extends beyond average Sharpe ratios to include downside protection and tail behavior. Models such as VLSTM, VxLSTM, and LPatchTST not only achieved strong full-sample Sharpe ratios, but also demonstrated superior Calmar ratios and materially higher minimum annual Sharpe ratios. In particular, LPatchTST exhibited an exceptionally stable worst-year profile, a property of practical relevance in institutional settings where drawdown control and avoidance of severely negative years are often prioritized over marginal improvements in average Sharpe.

Second, robustness must be evaluated multidimensionally. While VLSTM achieved the highest overall Sharpe ratio and strong passive-relative performance, VxLSTM and LPatchTST exhibited superior downside-adjusted characteristics in certain metrics, including Calmar ratio and minimum annual Sharpe. This distinction highlights an economically important trade-off: maximizing mean risk-adjusted return does not necessarily coincide with minimizing drawdown severity or tail exposure. The choice of architecture may therefore depend on investor preference over the mean–tail trade-off.

Third, turnover and transaction cost robustness meaningfully differentiate models. Breakeven transaction cost analysis reveals that xLSTM achieves the highest portfolio-level cost buffer, exceeding VLSTM despite slightly lower average Sharpe. This indicates greater resilience to implementation frictions and a higher signal-to-trade efficiency. Conversely, extremely low-turnover models such as iTransformer exhibit weak predictive performance, suggesting that insufficient responsiveness to evolving signals can be as detrimental as excessive trading. Successful architectures appear to strike a balance between adaptive state updating and controlled portfolio rebalancing.

Fourth, robustness to seed selection and experimental budget strengthens the credibility of the main findings. Performance rankings remain largely preserved under reduced seed aggregation, indicating that the reported differences are not artifacts of favorable initialization. This stability is particularly important in low signal-to-noise financial environments, where variance across runs can otherwise confound interpretation.

Fifth, state-space models and purely linear benchmarks exhibit heterogeneous and regime-sensitive behavior. While linear models occasionally perform well in high-volatility subperiods, they fail to deliver consistent cross-metric robustness. State-space architectures offer computational efficiency but require additional structure—such as feature selection or enriched state dynamics—to achieve competitive economic performance.

Taken together, the evidence suggests -- though does not definitively establish -- that successful financial forecasting architectures benefit from: (i) persistent and adaptively gated state representations, (ii) mechanisms for representation compression or feature conditioning, (iii) controlled trading intensity that preserves implementation robustness, and (iv) stability under adverse market realizations. Importantly, the most economically attractive models are not necessarily those with the highest average Sharpe, but those that jointly balance return, drawdown control, tail resilience, and transaction cost tolerance within realistic deployment constraints.

\subsection{Computational Efficiency and Model Complexity}

Table~\ref{tab:complexity} reported parameter counts as well as asymptotic runtime and memory complexities across models. The final number of trainable parameters and the corresponding hyperparameter ranges are provided in Appendix~\ref{appdx:hyperparams}. Linear models exhibited minimal computational costs in terms of both parameterization and asymptotic complexity, although their empirical performance in previous subsections was comparatively weaker within this benchmark.

Among nonlinear architectures, VLSTM combined competitive empirical performance with linear memory complexity in sequence length and without quadratic attention terms. Relative to transformer-based models such as PatchTST or TFT, this resulted in lower asymptotic memory growth with respect to sequence length. Mamba and Mamba2 offered comparable asymptotic efficiency, though their empirical performance in this study was more heterogeneous across evaluation metrics.

Hybrid architectures, including LPatchTST and VSN-enhanced variants, involved higher parameter counts and additional computational overhead. Within the present evaluation, these increases in complexity were not uniformly associated with proportional performance improvements. This pattern suggested that asymptotic complexity alone did not determine empirical effectiveness, and that architectural inductive bias may play an important role in financial time-series modeling. As with all preceding results, these observations are conditional on the specific dataset, sequence lengths, and implementation choices considered.






\begin{table*}[t]
\centering
\small
\begin{tabular}{lccc}
\hline
\textbf{Model} 
& \textbf{Number of Parameters} 
& \textbf{Runtime Complexity} 
& \textbf{Memory Complexity} \\
\hline

TFT
& $O(CH^2 + \ell H^2)$
& $O(L^2H + LCH^2)$
& $O(L^2 + LH)$ \\

VLSTM
& $O(CH^2 + H^2)$
& $O(LCH^2 + LH^2)$
& $O(LH)$ \\

Mamba
& $O(\ell H^2)$
& $O(LH^2)$
& $O(LH)$ \\

Mamba2
& $O(\ell H^2)$
& $O(LH^2)$
& $O(LH)$ \\

VSN + Mamba2
& $O(CH^2 + \ell H^2)$
& $O(LCH^2 + LH^2)$
& $O(LH)$ \\

AR1x
& $\mathbf{O(1)}$
& $\mathbf{O(L)}$
& $\mathbf{O(1)}$ \\

LSTM
& $\textcolor{gray}{O(H^2 + CH)}$
& $O(LH^2)$
& $O(LH)$ \\

NLinear
& $\underline{O(LC)}$
& $\underline{O(LC)}$
& $\underline{O(LC)}$ \\

DLinear
& $\underline{O(LC)}$
& $\underline{O(LC)}$
& $\underline{O(LC)}$ \\

xLSTM
& $\textcolor{gray}{O(H^2 + CH)}$
& $O(LH^2)$
& $O(LH)$ \\

Patch sLSTM
& $O(PCH + H^2)$
& $O(NH^2)$
& $\textcolor{gray}{O(NH)}$ \\

VSN + xLSTM
& $O(H^2(C + H))$
& $O(LH^2)$
& $O(LH^2)$ \\

PatchTST
& $O(PCH + \ell H^2)$
& $\textcolor{gray}{O(N^2H)}$
& $O(N^2 + NH)$ \\

LSTM + PatchTST
& $O(H^2 + PCH + \ell H^2)$
& $O(LH^2 + N^2H)$
& $O(LH + N^2)$ \\

iTransformer
& $O(\ell H^2)$
& $O(C^2H + LCH)$
& $O(C^2 + LC)$ \\

\hline
\end{tabular}
\caption{Comparison of parameter count, runtime complexity, and memory complexity for different time-series forecasting models. The emphasis on order of results is as follows: \textbf{best}, \underline{second\_best}, $\textcolor{gray}{third\_best}$. The forecasting horizon is fixed to one. $L$=sequence length, $C$=number of input features, $H$=hidden dimension, $N$=$L-S$=number of patches (for shifted versions adjusted for each time step), $P$=patch length, $S$=stride, $K$=convolution kernel size, $\ell$=number of layers, $M$=number of attention heads.}
\label{tab:complexity}
\end{table*}

%% file: Sections/conclusion.tex
This paper set out to answer a question motivated by the proliferation of deep learning architectures in time-series forecasting: how well do performance rankings from standard benchmarks translate to the challenging environment of financial asset returns? Using a large-scale benchmark of 15 models on a 15-year daily futures dataset, evaluated end-to-end on a volatility-targeted Sharpe ratio optimisation task, we provide a clear empirical answer.

\textbf{Answer to the title question.} The translation is poor. Several architectures that are widely considered 'state-of-the-art' on standard time-series benchmarks—including generic Transformers (iTransformer, PatchTST) and State Space Models (Mamba, Mamba2)—delivered out-of-sample Sharpe ratios below 1.0 in our financial forecasting task. Critically, these models often underperformed a simple AR(1)x benchmark, particularly in volatile regimes. This finding challenges the prevailing assumption that architectural advances validated on seasonal, high-signal benchmarks will automatically confer advantages in finance.

\textbf{Which architectures work, and why.} The most successful models in our benchmark were not the most complex or the most recent. Hybrid recurrent architectures—particularly VLSTM (VSN + LSTM), LPatchTST (LSTM + PatchTST), and TFT—achieved the highest and most stable risk-adjusted returns. xLSTM demonstrated superior transaction cost tolerance, while LPatchTST exhibited the strongest drawdown control. These results suggest that successful financial forecasting architectures share specific inductive biases: (i) persistent and adaptively gated state representations, (ii) mechanisms for representation compression or feature conditioning (e.g., Variable Selection Networks), and (iii) efficient translation of predictive signals into portfolio decisions. Generic attention mechanisms and unmodified SSMs lack these properties, which may explain their poor performance.

\textbf{Practical recommendations for forecasters.} Based on our findings, we offer the following guidance for researchers and practitioners applying deep learning to financial forecasting. These recommendations should be interpreted as empirical regularities observed under our experimental conditions (daily futures data, 2010–2025, volatility-targeted framework), not as universal guarantees.

\begin{itemize}
    \item \textbf{For maximising gross risk-adjusted returns:} Our results suggest that hybrid recurrent architectures—particularly VLSTM and LPatchTST—achieved the strongest Sharpe ratios. The common property shared by these models is the combination of explicit recurrent state representations with mechanisms for feature selection or temporal patching.
    \item \textbf{For robustness to transaction costs:} xLSTM demonstrated the highest breakeven cost buffer, primarily due to its lower turnover relative to other high-performing models. This suggests that architectures with enriched state representations may translate signals into portfolio decisions more efficiently.
    \item \textbf{For minimising drawdowns and tail risk:} LPatchTST exhibited the most stable worst-year performance and some of the strongest Calmar ratios, indicating that patch-based temporal aggregation may provide a useful inductive bias for controlling downside exposure.
    \item \textbf{What to avoid, cautiously:} Generic Transformers (iTransformer, vanilla PatchTST) and unmodified SSMs (Mamba, Mamba2) underperformed simpler methods in our benchmark. However, this finding is conditional on our specific implementation and dataset; modifications such as additional feature selection or recurrent pre-processing (as in LPatchTST) may alter performance.
\end{itemize}

Our conclusions are conditional on the specific dataset, asset classes (futures), sampling frequency (daily), and time period (2010–2025). Performance may differ on higher-frequency data (e.g., tick or minute bars), other asset classes (e.g., individual equities, cryptocurrencies, options), or different market regimes (e.g., prolonged bear markets, liquidity crises). Our transaction cost analysis assumes a simple proportional cost model, which may not capture market impact, slippage, or venue-specific frictions. Additionally, all models were evaluated under a volatility-targeting framework with a fixed 10\% target; different risk management protocols might alter relative rankings. Finally, while we used 50 random seeds and validated robustness with a reduced-seed experiment, the possibility of overfitting to the specific historical period cannot be entirely ruled out.

\textbf{Broader implications.} Our results suggest that the time-series forecasting community should be cautious about generalising benchmark results across domains. The statistical properties of the target series—signal-to-noise ratio, seasonality, non-stationarity, tail behaviour—are not ancillary details but central determinants of model performance. We recommend that future empirical studies in time-series forecasting explicitly report these properties and, where possible, evaluate models on multiple datasets with contrasting statistical characteristics. For the financial forecasting community, our benchmark provides a transparent, reproducible, and economically grounded reference point for future architectural development.

%% file: Sections/appendix.tex
\section{Data Construction and Exploratory Analysis}
\label{appdx:eda}

This appendix documents the construction of all variables used in the empirical analysis and provides detailed evidence on their distributional properties.

\subsection{Raw Data and Return Construction}

The raw dataset consists of daily observations with three fields: date, ticker, and closing price. From these, daily returns are computed as
\begin{equation}
r_t = \frac{P_t - P_{t-1}}{P_{t-1}},
\end{equation}
where $P_t$ denotes the closing price at time $t$.

\subsection{Volatility Estimation}

Daily volatility is estimated using an exponentially weighted moving average (EWMA) estimator. Let $\lambda = \frac{2}{\text{span}+1}$. The conditional mean and variance evolve according to
\begin{align}
\mu_t &= \lambda r_t + (1-\lambda)\mu_{t-1}, \\
\sigma_t^2 &= \lambda (r_t - \mu_t)^2 + (1-\lambda)\sigma_{t-1}^2.
\end{align}

\subsection{Distribution of Returns and Volatility}

\begin{figure}[t]
    \centering
    \includegraphics[width=\columnwidth]{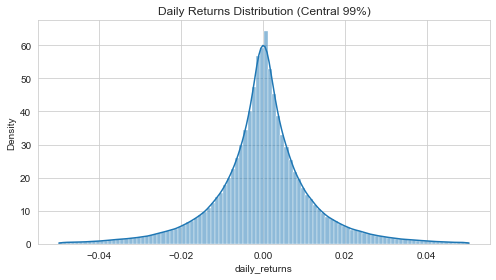}
    \caption{Distribution of daily returns. To make the central mass visible, the figure focuses on the bulk of the distribution; tail behavior is examined separately.}
    \label{fig:returns_dist}
\end{figure}

\begin{figure}[t]
    \centering
    \includegraphics[width=\columnwidth]{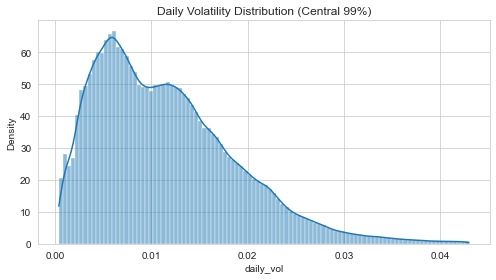}
    \caption{Distribution of daily realized volatility (log scale). Volatility exhibits strong right skewness and a long upper tail.}
    \label{fig:volatility_dist}
\end{figure}

Figure~\ref{fig:returns_dist} reports the empirical distribution of daily returns across all assets. The distribution is sharply peaked around zero and exhibits pronounced leptokurtosis. Figure~\ref{fig:volatility_dist} shows the distribution of realized volatility on a logarithmic scale, highlighting strong right skewness and a long upper tail.

\begin{figure*}[t]
    \centering
        \centering
        \includegraphics[width=\textwidth]{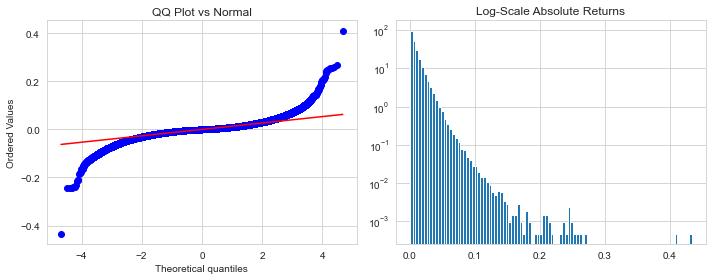}
        \caption{Left: Quantile--quantile plot against the normal distribution. Right: Tail behavior of daily returns. The figures indicate substantial deviations from Gaussianity and heavy-tailed return dynamics.}
    \label{fig:qq_logtails}
\end{figure*}

Figure~\ref{fig:qq_logtails} presents a quantile--quantile plot against the normal distribution together with the tail behavior of absolute returns. Both figures indicate substantial deviations from Gaussianity and slow tail decay.

\subsection{Predictive Features}

Normalized returns are constructed over multiple horizons (1 day, 1 week, 1 month, 3 months, 6 months, and 1 year) as
\begin{equation}
r^{\text{norm}}_{t,h} = \frac{r_{t,h}}{\sigma_t \sqrt{h}},
\end{equation}
where $h$ denotes the horizon in trading days.

Momentum indicators are further augmented using volatility-normalized and regime-adjusted \textit{Moving Average Convergence Divergence} (MACD) signals: 
\begin{align}
\operatorname{MACD}_t &= \operatorname{EWMA}_{h(T_s)}(P)_t - \operatorname{EWMA}_{h(T_l)}(P)_t, \\
q_t &= \frac{\operatorname{MACD}_t}{\operatorname{Std}_{63}(P)_t}, \\
\operatorname{Signal}_t &= \frac{q_t}{\operatorname{Std}_{252}(q)_t}.
\end{align}

\begin{figure*}[t]
    \centering
    \includegraphics[width=\textwidth]{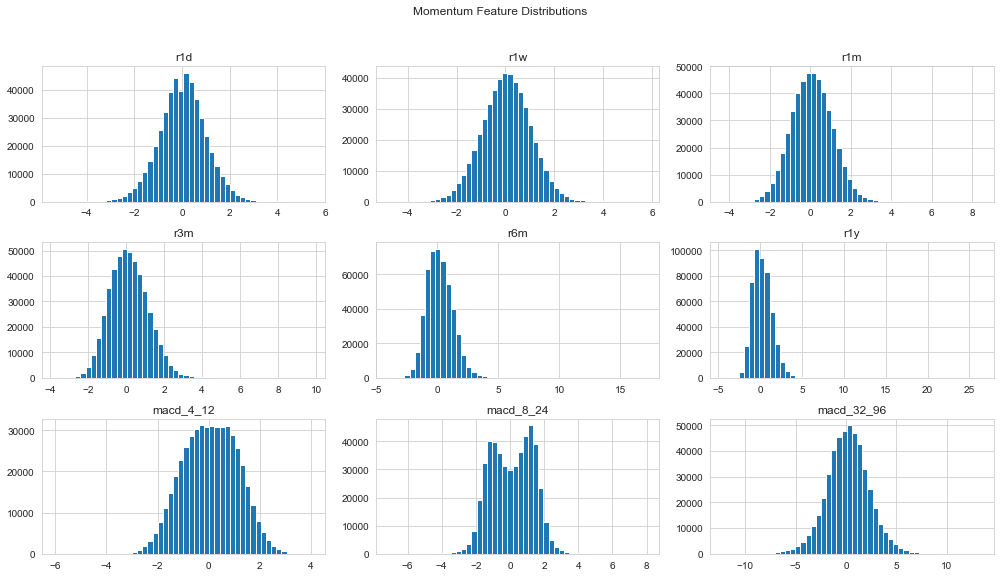}
    \caption{Distribution of Momentum Features.}
    \label{fig:data_mom}
\end{figure*}

The empirical distributions of these features are approximately symmetric and concentrated within the interval $[-2,2]$, with nearly all observations contained in $[-4,4]$, as shown in Figure~\ref{fig:data_mom}.

\subsection{Volatility-Scaled Exposure and Target Variable}

Volatility targeting induces a time-varying exposure factor defined as
\begin{equation}
\text{vs\_factor}_t = \frac{1}{\sigma_t}.
\end{equation}

\begin{figure}[t]
    \centering
    \includegraphics[width=\columnwidth]{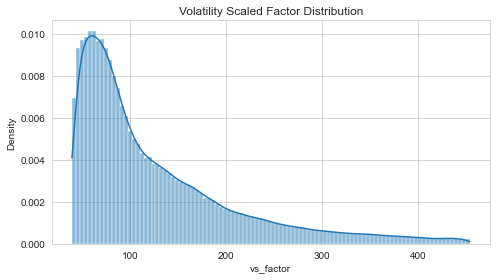}
    \caption{Induced Exposure by Volatility Targeting.
    }
    \label{fig:vs_factor}
\end{figure}

The empirical distribution of $vs\_factor$ exhibits pronounced right skewness, Figure~\ref{fig:vs_factor}. A large share of the mass is concentrated at relatively low values, while a long right tail corresponds to regimes or contracts characterized by unusually low realized volatility. These extreme realizations arise endogenously from the volatility-scaling mechanism itself and do not primarily reflect structural market dislocations or persistent risk premia.

From a modeling perspective, this feature is consequential. Linear specifications that treat volatility-scaled exposure as approximately proportional to changes in volatility may fail to capture the asymmetric response of leverage across regimes. Flexible nonlinear architectures are better suited to accommodate the threshold-like behavior induced by volatility targeting, especially in settings where exposure amplification during low-volatility periods magnifies the effect of predictive signals, while exposure compression during high-volatility episodes dampens their impact.

The learning target is constructed as the volatility-scaled next-period return,
\begin{equation}
\text{target}_t = \operatorname{clip}\left(\frac{r_{t+1}}{\sigma_t}, -20, 20\right),
\end{equation}
where clipping is applied to limit the influence of extreme realizations during training.

\subsection{Stylized Facts of Returns and Volatility}

Daily returns exhibit strong deviations from Gaussianity, with heavy tails and excess kurtosis that imply substantial downside risk and state dependence. Realized volatility is highly skewed and persistent, reflecting clustering and regime-dependent risk dynamics. These features challenge linear predictive models that rely on homoskedastic or symmetric error assumptions.

\subsection{Formal Statistical Diagnostics}

To formally substantiate the distributional characteristics suggested by the graphical evidence, we conduct standard diagnostic tests.

First, normality of returns is rejected at conventional significance levels using the Jarque–Bera test across the majority of instruments, consistent with the heavy tails observed in Figure~\ref{fig:qq_logtails}. Excess kurtosis and skewness statistics further confirm substantial deviations from Gaussianity.

Second, we test for non-stationarity in price levels using augmented Dickey–Fuller tests. As expected for financial price series, unit roots cannot generally be rejected in levels, while returns exhibit stationarity. This is consistent with the standard representation of asset prices as integrated processes with stationary increments.

Third, volatility persistence is assessed via autocorrelation of squared returns and realized volatility. We observe slow decay in the autocorrelation function, confirming volatility clustering and regime dependence.

Taken together, these diagnostics confirm that the dataset exhibits canonical stylized facts of financial time series: heavy tails, conditional heteroskedasticity, and non-stationarity in levels.

\subsection{Implications for Modeling}

Taken together, the distributional evidence highlights a fundamental feature of the data. While the input predictors are largely bounded, symmetric, and well-behaved, the target variable—realized returns—exhibits extreme kurtosis, heavy tails, and strong state dependence. This mismatch implies that the primary modeling challenge lies not in stabilizing the inputs, but in accurately capturing the conditional distribution of returns given these inputs.

These empirical properties motivate the use of flexible nonlinear architectures that can accommodate interactions, threshold effects, and regime-dependent behavior. Nonlinear dependencies and regime changes are frequently observed in financial time series and are captured most effectively by nonlinear models \cite{nonlinear_best}.  Neural network architectures such as LSTMs have shown superior performance in capturing complex nonlinear patterns in volatility and return dynamics \cite{nn_for_nonlinear}. In environments where small changes in predictors can, under certain conditions, lead to disproportionately large changes in outcomes, models that adapt their functional form across the state space are particularly well suited.

\subsection{Scope and External Validity}

While the dataset exhibits statistical properties commonly observed in financial markets, the conclusions drawn in this study are, strictly speaking, conditional on the data considered. As with any empirical investigation, it is not possible to establish universal generality beyond the sampled instruments and time period.

That said, the presence of heavy-tailed returns, volatility clustering, and regime-dependent dynamics suggests that the dataset captures structural characteristics typical of many financial time series. For this reason, we view the benchmark as representative of a broad class of cross-asset forecasting problems. Nevertheless, extending the analysis to alternative markets, frequencies, or macroeconomic environments remains an important direction for future research.

\section{Architecture Components}
\subsection{Linear Baselines}
\label{appdx:linear}
\paragraph{Autoregressive Model (AR1x).}
We consider an AR(1) process
\begin{equation}
    y_t = \phi y_{t-1} + \varepsilon_t, \quad \varepsilon_t \sim \mathcal{N}(0, \sigma^2),
\end{equation}
which captures short-term autocorrelation. This model assumes 1 one input feature. AR1x simply applies AR(1) to each feature independently (for multiple input features).

\paragraph{DLinear and NLinear.}
DLinear and NLinear \cite{nlinear} apply linear mappings to the input window:
\begin{equation}
    \hat{y}_t = W \mathbf{X}_t + b.
\end{equation}
DLinear decomposes the input into trend and seasonal components, while NLinear operates on normalized inputs. Both models lack temporal state and serve as non-recurrent linear baselines.

\subsection{Transformer Background}
\label{appdx:transformer}
Transformer architectures \cite{transformer} compute self-attention quadratically as
\begin{equation}
    \text{Attn}(Q,K,V) = \text{softmax}\!\left(\frac{QK^\top}{\sqrt{d_k}}\right)V.
\end{equation}
Therefore, Transformer-based models have a large number of trainable parameters.

\subsection{State-Space Model Details}
\label{appdx:mamba}
Mamba models \cite{mamba, mamba2} implement selective state-space models of the form
\begin{equation}
    h_t = A_t h_{t-1} + B_t x_t, \qquad y_t = C_t h_t,
\end{equation}
where $h_t$ denotes the latent state, $x_t$ the input, and $y_t$ the output. The matrices $A_t$, $B_t$, and $C_t$ are initialized using HiPPO LegS matrices \cite{hippo}, which provide a principled discretization of continuous-time linear dynamical systems.

In Mamba2, the state transition matrix is replaced with a scaled identity, and the architecture incorporates a form of linear attention \cite{lin_attn}, improving efficiency and numerical stability. 

In this work, we employ a static HiPPO transition matrix and a fixed low-rank parameterization of the step size $\Delta$, resulting in a fixed temporal horizon. This modification reduces sensitivity to noise and improves robustness when modeling financial time series.

\subsection{LSTM-based models}
\label{appdx:lstm}

The LSTM architecture is included as a canonical gated recurrent model that addresses the vanishing gradient limitations of standard RNNs through an additive memory cell and multiplicative gating. The forget and input gates enable adaptive control over memory retention and update, effectively implementing a data-dependent filtering mechanism.

The LSTM~\cite{lstm} updates are given by
\begin{align}
    f_t &= \sigma(W_f x_t + U_f h_{t-1} + b_f), \\
    i_t &= \sigma(W_i x_t + U_i h_{t-1} + b_i), \\
    o_t &= \sigma(W_o x_t + U_o h_{t-1} + b_o), \\
    c_t &= f_t \odot c_{t-1} + i_t \odot \tanh(W_c x_t), \\
    h_t &= o_t \odot \tanh(c_t),
\end{align}
where $f_t$, $i_t$, and $o_t$ denote the forget, input, and output gates, respectively. Such adaptive memory is particularly relevant in financial time series, where structural breaks and regime shifts render fixed-memory models suboptimal. The gating mechanism allows the model to dynamically adjust the effective time horizon over which past information is retained.

Moreover, the nonlinear hidden-state representation provides a flexible mechanism for extracting predictive structure from low signal-to-noise and non-Gaussian data, making LSTMs a natural and widely adopted baseline in financial forecasting tasks.

\paragraph{xLSTM.}

We provide a concise technical description of xLSTM following \cite{xlstm} (Eq.~8–17).

\subparagraph{Exponential Gating.}

In contrast to classical LSTMs, which employ sigmoid gates $f_t = \sigma(\cdot)$ and $i_t = \sigma(\cdot)$, xLSTM replaces sigmoid activations with exponential gating followed by normalization. Let

\begin{align}
\tilde{f}_t &= W_f x_t + R_f h_{t-1} + b_f, \\
\tilde{i}_t &= W_i x_t + R_i h_{t-1} + b_i,
\end{align}

which are the pre-activation gates, similar to LSTM. The raw gates are exponentiated,

\begin{equation}
\hat{f}_t = \exp(\tilde{f}_t), 
\qquad
\hat{i}_t = \exp(\tilde{i}_t),
\end{equation}

and subsequently normalized to ensure numerical stability and controlled memory growth. In log-domain form, this normalization is implemented via a running maximum term to prevent overflow (see \cite{xlstm}, Eq.~15–17). The resulting gates satisfy a convex combination structure analogous to classical LSTM gating but without sigmoid saturation.

\subparagraph{sLSTM (Scalar LSTM).}

The scalar variant maintains a single memory state per unit. The update equations are

\begin{align}
\tilde{c}_t &= \tanh(W_z x_t + R_z h_{t-1} + b_z), \\
c_t &= f_t \odot c_{t-1} + i_t \odot \tilde{c}_t, \\
n_t &= f_t \odot n_{t-1} + i_t, \\
h_t &= o_t \odot \frac{c_t}{n_t},
\end{align}

where $f_t$, $i_t$, and $o_t$ denote normalized exponential forget, input, and output gates, respectively, and $n_t$ is a stabilizing normalizer. The division by $n_t$ ensures scale control of the memory state.

\subparagraph{mLSTM (Matrix LSTM).}

The matrix variant generalizes the memory to a matrix-valued state $C_t \in \mathbb{R}^{d \times d}$. At each step, key-value vectors $k_t, v_t \in \mathbb{R}^d$ are generated and stored via a gated outer-product update:

\begin{equation}
C_t = f_t \odot C_{t-1} + i_t \odot (v_t k_t^\top).
\end{equation}

The hidden state is retrieved through a query vector $q_t$,

\begin{equation}
h_t = C_t q_t,
\end{equation}

This Bidirectional Associative Memory (BAM) \cite{BAM1,BAM2,BAM3,BAM4} setting yields high separability between stored patterns and allows efficient recall of past information. Importantly, mLSTM removes state compression recurrence but still has temporal dependency in memory accumulation. It is more parallelizable, but not fully parallel.

\paragraph{Patch sLSTM (PsLSTM).}

Given a multivariate time series $X \in \mathbb{R}^{L \times d}$, each channel is treated independently and segmented into non-overlapping patches of length $\ell$:

\begin{equation}
\tilde{x}_p^{(i)} = \mathrm{Patch}\left(x_{(p-1)\ell+1:p\ell}^{(i)}\right),
\quad i = 1,\dots,d.
\end{equation}

Each patch embedding is processed by an sLSTM with shared parameters across channels:

\begin{equation}
h_p^{(i)} = \mathrm{sLSTM}(h_{p-1}^{(i)}, \tilde{x}_p^{(i)}).
\end{equation}

Parameter sharing preserves channel independence while reducing model complexity and mitigating overfitting.

\subsection{Hybrid Architecture Details}
\label{appdx:hybrid}

The hybrids are designed to improve robustness in noisy financial time series by enhancing the signal-to-noise ratio, enabling adaptive feature selection, and stabilizing temporal state updates.

\paragraph{Variable Selection Network (VSN)}
\label{appdx:vsn}

The Variable Selection Network (VSN), inspired by the Temporal Fusion Transformer \cite{tft}, performs feature-wise nonlinear embedding followed by dynamic soft selection of relevant covariates at each time step.

Given an input vector $x_t \in \mathbb{R}^{C}$ consisting of $C$ covariates, each variable is embedded independently:
\begin{equation}
    h_{t,i} = \phi_i(x_{t,i}), \quad i = 1,\dots,C,
\end{equation}
where $\phi_i(\cdot)$ denotes a learnable nonlinear embedding function.

The embeddings are concatenated and passed through a gating network to compute feature importance weights:
\begin{equation}
    \alpha_t = \text{softmax}\left(W_g [h_{t,1}, \dots, h_{t,C}] + b_g \right),
\end{equation}
where $W_g$ and $b_g$ are learnable parameters.

The selected representation is computed as a weighted sum:
\begin{equation}
    \tilde{x}_t = \sum_{i=1}^{C} \alpha_{t,i} h_{t,i}.
\end{equation}

This mechanism enables adaptive suppression of noisy or uninformative covariates and improves robustness in non-stationary environments.

\paragraph{VSN+LSTM (VLSTM)}

The VSN+LSTM (VLSTM) model combines VSN-based feature selection with recurrent temporal encoding. At each time step, the input vector is processed by a VSN to produce a dynamically weighted feature representation:
\begin{equation}
    \tilde{x}_t = \mathrm{VSN}(x_t).
\end{equation}

The resulting sequence $\{\tilde{x}_t\}_{t=1}^{L}$ is then passed through an LSTM to construct a compact temporal state:
\begin{equation}
    (h_t, c_t) = \mathrm{LSTM}(\tilde{x}_t, h_{t-1}, c_{t-1}).
\end{equation}

For one-step-ahead forecasting, the prediction is obtained from the final hidden state:
\begin{equation}
    \hat{y}_{L+1} = W_o h_L + b_o.
\end{equation}

\paragraph{VSN--Mamba2}

In the VSN--Mamba2 hybrid, feature selection and temporal modeling are explicitly decoupled. Given an input $x_t \in \mathbb{R}^{d}$, the VSN computes feature-wise importance weights:
\begin{equation}
    \alpha_t = \text{softmax}(g(x_t)), \qquad \tilde{x}_t = \alpha_t \odot x_t,
\end{equation}
where $g(\cdot)$ denotes a learnable gating network and $\odot$ denotes element-wise multiplication.

The filtered input $\tilde{x}_t$ is then passed to the Mamba2 state-space block:
\begin{equation}
    h_t = A h_{t-1} + B \tilde{x}_t,
\end{equation}
where $A$ and $B$ denote the state transition and input matrices, respectively. This design improves robustness by reducing the influence of noisy covariates prior to temporal state updates.

\paragraph{LSTM + PatchTST}

This hybrid architecture combines explicit recurrence with attention by using an LSTM as a channel-wise temporal denoiser prior to PatchTST. Each input channel is processed independently using a shared LSTM backbone:
\begin{equation}
    h_t^{(i)} = \mathrm{LSTM}(x_t^{(i)}, h_{t-1}^{(i)}), \quad i = 1,\dots,d,
\end{equation}
where $x_t^{(i)}$ denotes the $i$-th feature.

The resulting hidden states are segmented into temporal patches and passed to PatchTST, which applies self-attention over patches:
\begin{equation}
    \tilde{h}_p = \mathrm{PatchTST}(\{h_t^{(i)}\}_{t \in \mathcal{P}_p}),
\end{equation}
where $\mathcal{P}_p$ denotes the set of time steps belonging to patch $p$.

This separation of concerns allows the LSTM to stabilize local temporal structure while PatchTST aggregates medium- and long-range dependencies.

\paragraph{VSN + xLSTM}

In the VSN + xLSTM hybrid, the sequence of VSN-selected representations is directly fed into an xLSTM. The model maintains a matrix-valued memory state $M_t \in \mathbb{R}^{H \times H}$, which is updated recursively:
\begin{equation}
    (M_t, s_t) = \mathrm{xLSTM}(\tilde{x}_t, M_{t-1}),
\end{equation}
where $s_t \in \mathbb{R}^{H}$ denotes the output state.
The matrix-valued memory enables modeling of higher-order temporal interactions and long-range dependencies beyond the capacity of vector-based recurrent architectures.

\subsection{Temporal Fusion Transformer Details}

TFT combines recurrent encoding with attention-based aggregation. Given an input sequence $\{x_t\}_{t=1}^L$, a recurrent encoder produces latent states
\[
h_t = \mathrm{LSTM}(x_t, h_{t-1}).
\]
Multi-head attention is then applied to aggregate information across time:
\[
\hat{y}_t = \mathrm{Attn}(h_{1:t}).
\]
Variable selection networks, gating layers, and static covariate encoders further modulate the representations to improve interpretability and robustness.

\section{Model Configurations and Hyperparameter Selection}
\label{appdx:hyperparams}

To ensure a fair and systematic comparison across architectures, we define structured hyperparameter search spaces tailored to each model family. All models are evaluated under comparable training budgets, with architecture-specific parameters varied only where structurally relevant. Tables~\ref{tab:model_overview} and~\ref{tab:arch_specific} summarizes the general configuration ranges.

\paragraph{General Training Configuration.}
For each model, we tune the batch size, hidden dimensionality ($H$), learning rate, and input sequence length. Learning rates are selected from logarithmic ranges between $10^{-4}$ and $10^{-2}$ or from discrete sets $\{10^{-3}, 5\times10^{-4}, 10^{-4}\}$ depending on architectural stability. Sequence lengths are chosen to reflect short- and medium-horizon dependencies (e.g., 64--512 timesteps), while maintaining comparability across models.

\paragraph{Architecture-Specific Hyperparameters.}
In addition to shared training parameters, each model family includes structural hyperparameters:

\begin{itemize}
    \item \textbf{Transformer-based models} (e.g., TFT, PatchTST) vary the number of attention heads, encoder layers, and sparsity factors.
    \item \textbf{State-space models} (Mamba2 variants) tune the number of layers, convolution kernel size, SSM expansion factors, rank, and discretization parameters (e.g., $\Delta_{\max}$, HiPPO scaling).
    \item \textbf{xLSTM-based models} vary the number of stacked blocks, convolutional kernel sizes, and projection expansion factors.
    \item \textbf{Patch-based models} additionally tune patch length and stride fraction.
    \item \textbf{Classical baselines} (AR1, DLinear) are evaluated with minimal architectural tuning.
\end{itemize}

\paragraph{Search Protocol.}
Hyperparameter selection is performed via grid search over the predefined discrete ranges. Each configuration is trained independently, and the best-performing setting is selected based on validation Sharpe Ratio. Importantly, the same validation procedure and performance metric are used for all architectures to avoid selection bias.
\paragraph{Training Details and Evaluation Alignment.}
We employ the ADAM optimizer~\cite{adam} for all models. Early stopping is applied with a patience of 20 epochs. For validation, the last 10\,\% of the training data is reserved as a validation set. Gradient clipping based on the gradient norm is used to stabilize training.

Models are retrained every five years using a rolling-window scheme. For strategies that require an initial lookback period, performance metrics are not evaluated during the burn-in phase. The burn-in length is set to leave 21 days for evaluation (63 days for sequence length 84 and 491 days for sequence length 512). Critically, the calendar start date of the evaluation window is identical for every architecture: after discarding its own burn-in, each model begins scoring on the same first trading day. Consequently, all reported metrics are computed on exactly the same set of out-of-sample days, so differences in performance cannot be attributed to mismatched evaluation periods.

\paragraph{Model Capacity Considerations.}
The number of trainable parameters varies substantially across models (Table~\ref{tab:model_overview}), reflecting inherent architectural differences. We do not explicitly match parameter counts in the main experiments, as forcing an identical budget would distort the native design of certain architectures (e.g., compact state-space models versus heavily parameterized attention- or recurrence-based models). Instead, every architecture is given access to the same broad hyperparameter search space; the final configuration reported in Table~\ref{tab:model_overview} is the one that maximised validation performance for that architecture. Consequently, differences in parameter count are themselves an outcome of the search rather than an arbitrary choice.

To verify that the ranking is not an artefact of uncontrolled capacity, we re-ran the full evaluation under a hard constraint that every model must use between 1\,M and 2\,M trainable parameters. Hyperparameters were re-selected inside this band (or left unchanged when the original optimum already lay inside it). Table~\ref{tab:capacity_controlled} reports the resulting Sharpe ratios and key secondary metrics. The relative ordering is essentially unchanged: VLSTM remains the strongest model (its original parameter count already lay inside the admissible range), followed by the same group of gated-recurrent and hybrid architectures. Models that originally operated with far fewer or far more parameters show inferior performances as expected. These results indicate that the performance gaps observed in the main text are driven primarily by architectural inductive bias rather than by differences in raw capacity.

\begin{table*}[t]
\small
\setlength{\tabcolsep}{4pt} 
\centering
\caption{Model configurations and general hyperparameter search ranges. 
}
\begin{adjustbox}{width=\textwidth} 
\begin{tabular}{lcccccc}
\toprule
Model & Trainable Params & Batch Size & Hidden Dim (H) & LR Range & Seq Len & Dropout \\
\midrule
Mamba2+VSN      & 3,571,348 & \{32,64,128\} & \{32,64,128,256\} & $10^{-4}$--$10^{-2}$ & 84 & \{0.1,0.2,0.3,0.4\} \\
xLSTM           & 2,507,269 & \{128,256\}   & \{64,128,256\}    & \{1e-3,5e-4,1e-4\} & 84 & \{0.2,0.3,0.4,0.5\} \\
xLSTM+VSN       & 6,146,368 & \{128,256\}   & \{64,128,256\}    & \{1e-3,5e-4,1e-4\} & 84 & \{0.2,0.3,0.4,0.5\} \\
PsLSTM          & 1,963,841 & \{16,32\}     & \{16,32,64\}      & \{1e-3,5e-4,1e-4\} & 512 & \{0.2,0.3,0.4,0.5\} \\
VLSTM           & 1,142,963 & \{128,256\}   & \{64,128,256\}    & $10^{-4}$--$10^{-2}$ & 84 & \{0.2,0.3,0.4,0.5\} \\
LSTM            & 73,729    & \{128,256\}   & \{64,128,256\}    & $10^{-4}$--$10^{-2}$ & 84 & \{0.2,0.3,0.4,0.5\} \\
PatchTST        & 1,139,739 & \{16,32\}     & \{16,32,64,96\}   & \{1e-3,5e-4\} & 512 & \{0.1,0.2,0.3,0.4,0.5\} \\
LSTM+PatchTST   & 634,881   & \{16,32\}     & \{16,32,64,96\}   & \{1e-3,5e-4\} & 512 & \{0.1,0.2,0.3,0.4,0.5\} \\
Mamba2          & 18,882    & \{32,64,128\} & \{32,64,128,256\} & $10^{-4}$--$10^{-2}$ & 84 & \{0.1,0.2,0.3,0.4\} \\
DLinear         & 2,111     & \{256,512,1024\}  & \{64,128,256\}    & $10^{-4}$--$10^{-2}$ & 84 & \{0.1,0.2,0.3\} \\
AR1x             & 2,073     & \{128,256\}   & \{64,128,256\}    & $10^{-4}$--$10^{-2}$ & 84 & \{0.2,0.3,0.4,0.5\} \\
TFT             & 347,507   & \{128,256\}   & \{64,128,256\}    & $10^{-4}$--$10^{-2}$ & 84 & \{0.2,0.3,0.4,0.5\} \\
iTransformer             & 1,007,505   & \{64,128\}   & \{64,128,256,512\}    & $10^{-4}$--$10^{-3}$ & 84 & \{0.1,0.2,0.3,0.4,0.5\} \\
\bottomrule
\end{tabular}
\end{adjustbox}
\label{tab:model_overview}
\end{table*}

\begin{table*}[t]
\centering
\caption{Architecture-specific hyperparameter search spaces.}
\label{tab:arch_specific}
\begin{adjustbox}{width=\textwidth}  
\begin{tabular}{ll}
\toprule
Architecture & Hyperparameter Search Space \\
\midrule

Transformer-based 
& Heads: \{1,2,4\}; Layers: \{3,4,5,6\}; Sparsity: \{2,3,4,5,6\} \\

Mamba / SSM 
& Layers: \{1,2,3,5\}; Kernel: \{2,3,5,7,9\}; Conv: \{2,8,16,32\} \\
& Rank: \{4,8,16,32\}; $\Delta_{\max}$: \{0.2,0.4,0.6,0.8\}; HiPPO: 0.1--0.5 \\

xLSTM 
& Blocks: \{1--6\}; Kernel: \{1,2,4,6,7,9\}; Projection: \{1--2.5\} \\

Patch-based 
& Patch Length: \{4,8,16,32,64\}; Stride: \{0.25,0.5,1\} \\

Classical (AR1x, DLinear)
& Layers: \{2,3,5\} \\

\bottomrule
\end{tabular}
\end{adjustbox}
\end{table*}

\begin{table*}[t]
\centering
\caption{Performance under a controlled capacity budget (1--2\,M trainable parameters).}
\label{tab:capacity_controlled}
\begin{tabular}{lccccccccc}
\toprule
Model & Params & SR & CAGR & Max DD & t (HAC) & Calmar & Turnover & Info. Ratio & Hit \\
\midrule
VLSTM           & 1,142,963 & \textbf{2.40} & \textbf{0.2632} & -0.229 & \textbf{8.81} & 1.15 & 966.86 & \textbf{0.854} & \textbf{0.588} \\
PsLSTM          & 1,963,841 & 1.78 & 0.1923 & \underline{-0.127} & \underline{7.17} & \textbf{1.51} & 884.61 & 0.389 & 0.560 \\
PatchTST        & 1,139,739 & 0.75 & 0.0855 & -0.184 & 3.31 & 0.47 & 613.85 & -0.199 & 0.538 \\
LPatchTST       & 1,714,433 & 0.56 & 0.0393 & -0.325 & 1.53 & 0.12 & \underline{105.88} & -0.176 & 0.532 \\
Mamba2          & 1,967,714 & 0.46 & 0.0376 & -0.277 & 1.55 & 0.14 & 138.83 & -0.161 & 0.532 \\
VMamba2         & 1,096,304 & 1.02 & 0.1617 & \textbf{-0.116} & 5.85 & \underline{1.39} & 549.06 & 0.505 & 0.564 \\
xLSTM           & 1,047,050 & 1.70 & 0.1709 & -0.136 & 6.00 & 1.26 & 372.79 & 0.797 & 0.561 \\
VxLSTM          & 1,086,306 & 1.68 & 0.1907 & -0.141 & 6.64 & 1.35 & 482.62 & \underline{0.798} & 0.568 \\
LSTM            & 1,081,345 & 1.22 & 0.124 & -0.367 & 4.18 & 0.34 & 992.09 & -0.177 & 0.557 \\
TFT             & 1,341,875 & \underline{2.05} & \underline{0.205} & -0.291 & 6.97 & 0.70 & 877.61 & 0.443 & \underline{0.576} \\
iTransformer    & 1,007,505 & 0.38 & 0.0308 & -0259 & 1.26 & 0.12 & \textbf{36.32} & -0.154 & 0.529 \\
\bottomrule
\end{tabular}
\end{table*}

\section{Performance Metrics and Evaluation Criteria}
\label{appdx:metrics}

This appendix describes the performance, risk, and robustness measures used throughout the empirical evaluation. Given the low signal-to-noise ratio and heavy-tailed nature of financial returns, we rely on a broad set of complementary metrics to assess not only average performance, but also statistical significance, downside risk, trading intensity, and incremental value relative to a passive benchmark.

\subsection{Return and Risk-Adjusted Performance}
\label{appdx:measures}
\paragraph{Annualized Return.}
Annualized return is computed as the mean daily portfolio return scaled by the number of trading days per year. While intuitive, this metric does not account for risk and is therefore interpreted jointly with risk-adjusted measures.

\paragraph{Compound Annual Growth Rate (CAGR).}
CAGR measures the geometric average annual growth of portfolio value over the evaluation period:
\[
\text{CAGR} = \left( \frac{V_T}{V_0} \right)^{1/T} - 1,
\]
where \(V_0\) and \(V_T\) denote initial and terminal portfolio values and \(T\) is the length of the sample in years. CAGR reflects long-run capital accumulation and penalizes volatility drag.

\paragraph{Sharpe Ratio.}
The Sharpe ratio is defined as the mean excess return divided by the standard deviation of returns. All reported Sharpe ratios are annualized. This is the primary optimization objective and headline performance metric throughout the study. Note that for future contracts (which we consider here) there is no need to borrow. Thus we can consider the risk-free-rate to be zero (reflected in equation~\ref{eq:sr}).

\paragraph{Information Ratio.}
The Information Ratio measures risk-adjusted excess performance relative to a passive buy-and-hold benchmark:
\[
\text{IR} = \frac{\hat{\mathbb{E}}[r_t - r_t^{\text{passive}}]}{\sqrt{\hat{\text{Var}}(r_t - r_t^{\text{passive}})}}.
\]
This metric captures incremental value beyond market exposure.

\subsection{Statistical Significance}
\label{appdx:stat_signif} 
\paragraph{HAC-Adjusted \(t\)-Statistics.}
To assess statistical significance under serial correlation and heteroskedasticity, we report Newey--West heteroskedasticity and autocorrelation consistent (HAC) \(t\)-statistics for mean returns and Sharpe ratios. This adjustment is critical in financial time series, where returns often exhibit time dependence.

\paragraph{HAC \(t\)-Statistic versus Passive.}
We additionally report HAC-adjusted \(t\)-statistics for excess returns relative to the passive benchmark, testing whether observed outperformance is statistically distinguishable from zero after accounting for dependence in relative returns.

\subsection{Directional Accuracy and Trading Activity}

\paragraph{Hit Rate.}
The hit rate measures the fraction of periods in which the strategy correctly predicts the sign of returns. While not sufficient for profitability on its own, it provides insight into directional consistency.

\paragraph{Turnover.}
Turnover is defined as the average absolute change in portfolio weights across consecutive periods. High turnover implies greater transaction costs and reduced implementability.

\paragraph{Turnover (xGMV).}
Turnover expressed as a multiple of gross market value (xGMV) provides a scale-free measure of trading intensity and facilitates comparison across strategies.

\subsection{Downside Risk and Tail Behavior}

\paragraph{Maximum Drawdown.}
Maximum drawdown is the largest peak-to-trough decline in cumulative portfolio value. It captures worst-case capital loss and is a key risk metric for real-world deployment.

\paragraph{Calmar Ratio.}
The Calmar ratio is defined as CAGR divided by maximum drawdown. It measures return efficiency relative to extreme downside risk and complements the Sharpe ratio.

\paragraph{Conditional Value-at-Risk (CVaR 5\%).}
CVaR at the 5\% level measures the expected loss conditional on returns falling in the worst 5\% of outcomes. Unlike Value-at-Risk, CVaR captures tail severity and is particularly relevant in heavy-tailed financial return distributions.

\subsection{Benchmark-Relative Diagnostics}

\paragraph{Correlation versus Passive.}
We report the Pearson correlation between strategy returns and the passive benchmark. Lower correlation indicates greater diversification benefits and reduced dependence on market direction.

\paragraph{Profit and Loss (PnL).}
Cumulative profit and loss (PnL) curves are used for visual comparison of strategies over time. PnL trajectories provide insight into path dependence, drawdown behavior, and regime sensitivity that may not be evident from summary statistics alone.

\medskip
Taken together, these metrics provide a comprehensive and economically meaningful assessment of model performance, balancing average returns, statistical reliability, downside risk, trading realism, and incremental value over a passive investment strategy.

\section{Asset-level Results}
\label{appdx:asset_level}
This section provides a detailed breakdown of model performance at the individual asset level. The per-asset analysis presented here serves two purposes: (i) to evaluate cross-sectional robustness and (ii) to assess whether performance is concentrated in specific asset classes or broadly distributed across markets.

We use \textit{breakeven transaction cost}, which is defined as the constant cost per unit of leveraged turnover that drives total gross PnL to zero. Formally,
\[
c^{*} = \frac{\sum_{t} R^{\text{gross}}_{t}}{\sum_{t} \tau_{t}},
\]
where $\tau_{t}$ denotes leveraged turnover. If actual trading costs remain below $c^{*}$, the strategy remains profitable; the ratio $c^{*}/c_{\text{actual}}$ measures implementation robustness.

Results are grouped by asset category (Bond, Commodity, Energy, FX, and Index futures). For each model, we report cumulative PnL trajectories per asset as well as box-and-whisker summaries of the Sharpe Ratio distributions. This allows us to examine both temporal consistency and cross-sectional dispersion.

\subsection{Asset Description}
Table~\ref{tbl:asset_universe} reports the full universe of tradable assets included in the empirical analysis. The cross-asset dataset spans foreign exchange, equity indices, fixed income, energy, and agricultural and metal commodities, including both pit and electronic contracts where available. This broad coverage ensures substantial cross-sectional and cross-asset heterogeneity, enabling the evaluation of model robustness across diverse liquidity conditions, macroeconomic exposures, and volatility regimes.

\begin{table*}[t]
\centering
\scriptsize
\setlength{\tabcolsep}{3pt}
\caption{Asset universe and classification.}
\begin{tabular}{lll | lll | lll}
\hline
Ticker & Group & Description &
Ticker & Group & Description &
Ticker & Group & Description \\
\hline
AN & FX & Australian dollar comp &
BC & Energy & Brent crude oil &
BG & Energy & Brent gasoil  \\

BN & FX & British pound comp &
CA & Index & CAC 40 index &
CADJPY & FX & CAD JPY cross \\

CB & Bond & Canada 10Y bond &
CC & Comdty & Cocoa &
CN & FX & Canadian dollar comp \\

CR & Comdty & CRB index &
CT & Comdty & Cotton No.2 &
DA & Comdty & Milk III \\

DT & Bond & Euro Bund &
DX & FX & US dollar index &
EN & Index & Nasdaq mini \\

ER & Index & Russell 2000 mini &
ES & Index & S\&P 500 mini &
FB & Bond & US 5Y note comp \\

FN & Index & Euro Stoxx comp &
GI & Comdty & Goldman Sachs idx &
GS & Bond & UK Gilt long \\

HS & Index & Hang Seng &
JN & FX & Japanese yen comp &
JO & Comdty & Orange juice \\

KC & Comdty & Coffee &
KW & Comdty & Wheat KC &
LB & Comdty & Lumber \\

LX & Index & FTSE 100 &
MD & Index & S\&P 400 mini &
MP & FX & Mexican peso \\

MW & Comdty & Wheat Minneapolis &
NK & Index & Nikkei 225 &
NOKUSD & FX & NOK USD cross \\

NR & Comdty & Rough rice &
SB & Comdty & Sugar No.11 &
SN & FX & Swiss franc comp \\

TU & Bond & US 2Y note comp &
TY & Bond & US 10Y note comp &
UB & Bond & Euro Bobl \\

US & Bond & US T-bond comp &
USDNZD & FX & USD NZD cross &
USDSEK & FX & USD SEK cross \\

USDSGD & FX & USD SGD cross &
UZ & Bond & Euro Schatz &
XU & Index & Euro Stoxx 50 \\

XX & Index & STOXX 50 &
YM & Index & Mini Dow &
ZA & Comdty & Palladium elec \\

ZB & Energy & RBOB elec &
ZC & Comdty & Corn elec &
ZF & Comdty & Feeder cattle elec \\

ZG & Comdty & Gold elec &
ZI & Comdty & Silver elec &
ZK & Comdty & Copper elec \\

ZL & Comdty & Soybean oil elec &
ZM & Comdty & Soybean meal elec &
ZN & Energy & Natural gas elec \\

ZO & Comdty & Oats elec &
ZP & Comdty & Platinum elec &
ZR & Comdty & Rough rice elec \\

ZS & Comdty & Soybeans elec &
ZT & Comdty & Live cattle elec &
ZU & Energy & Crude oil elec \\

ZW & Comdty & Wheat elec &
ZZ & Comdty & Lean hogs elec & \\

\hline
\end{tabular}
\label{tbl:asset_universe}
\end{table*}

\subsection{VLSTM}
\begin{figure*}[h]
    \centering
    \includegraphics[width=0.8\textwidth]{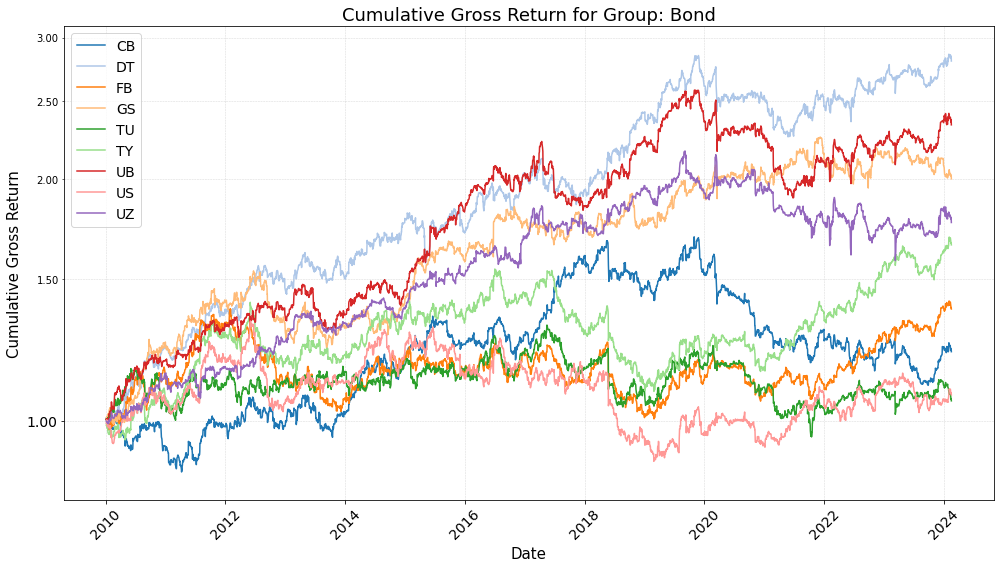}
    \caption{VLSTM PnL per asset - bond Futures} 
    \label{fig:vlstm_asset_bond}
\end{figure*}

\begin{figure*}[h]
    \centering
    \includegraphics[width=0.8\textwidth]{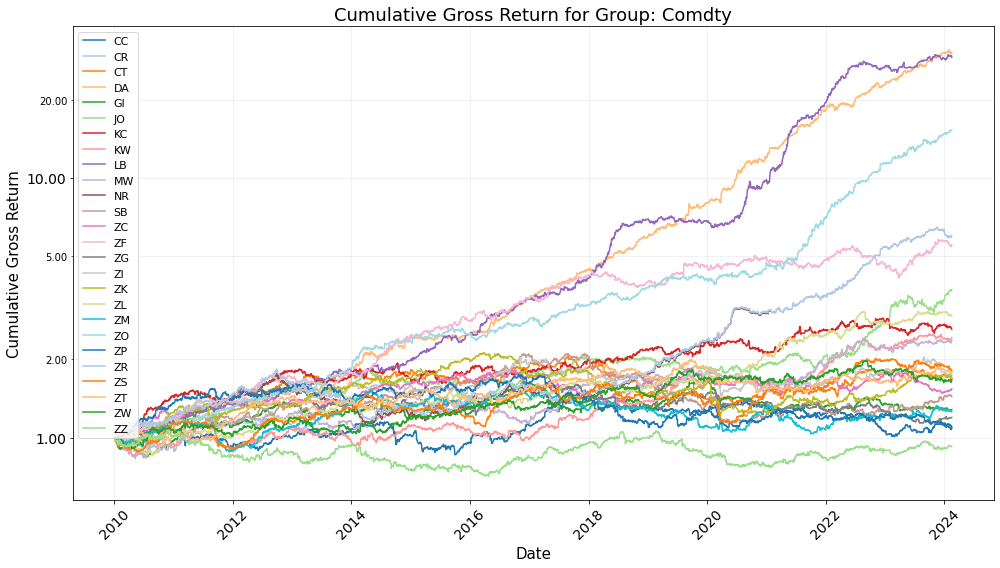}
    \caption{VLSTM PnL per asset - Commodities Futures} 
    \label{fig:vlstm_asset_cmdty}
\end{figure*}

\begin{figure*}[h]
    \centering
    \includegraphics[width=0.8\textwidth]{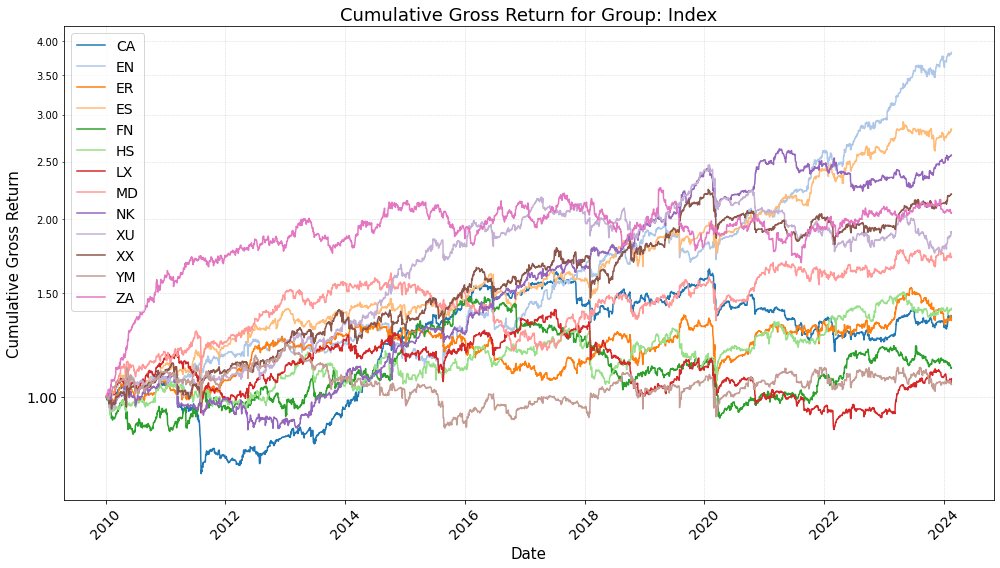}
    \caption{VLSTM PnL per asset - Index Futures}
    \label{fig:vlstm_asset_index}
\end{figure*}

\begin{figure*}[h]
    \centering
    \includegraphics[width=0.8\textwidth]{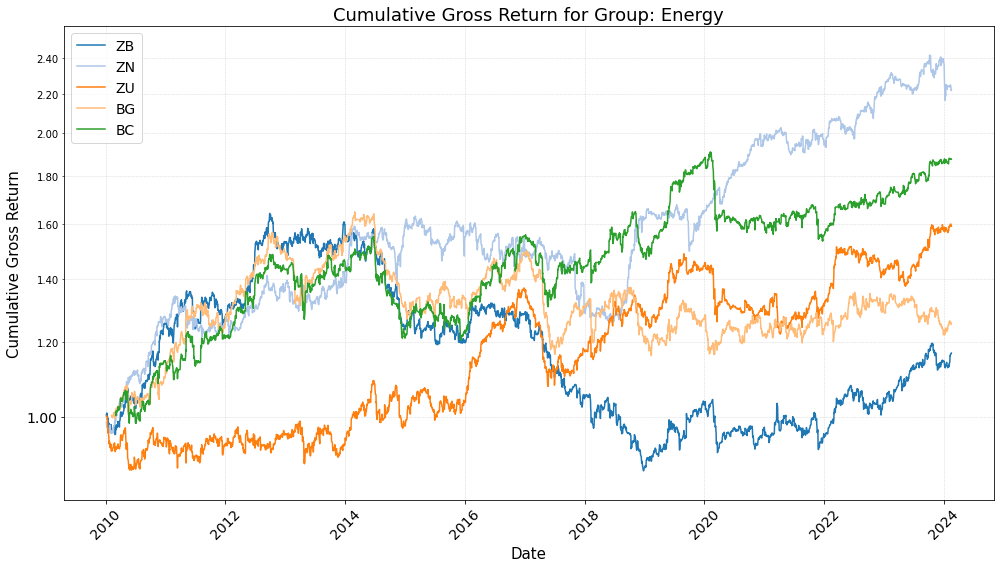}
    \caption{VLSTM PnL per asset - Energy Futures}
    \label{fig:vlstm_asset_energy}
\end{figure*}

\begin{figure*}[h]
    \centering
    \includegraphics[width=0.8\textwidth]{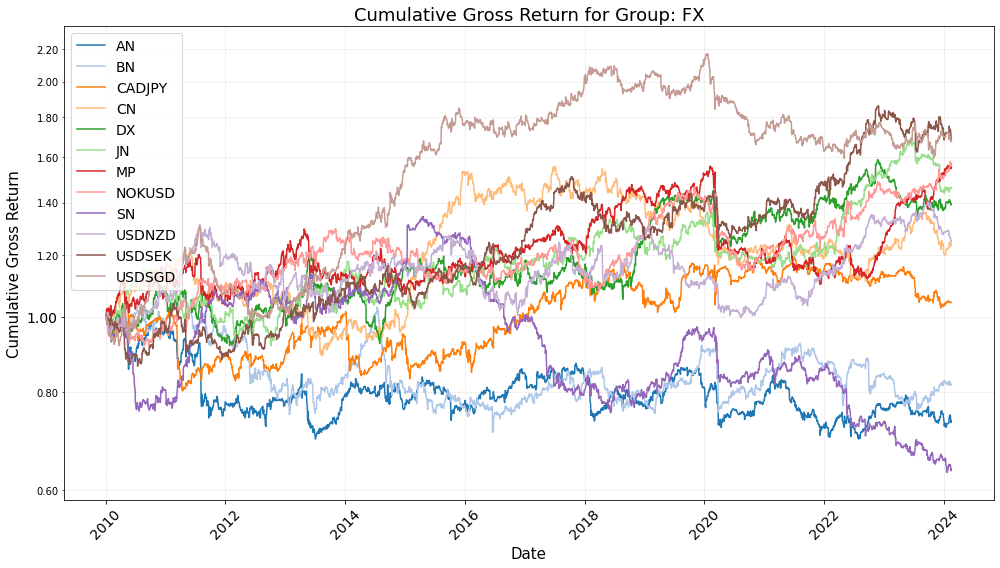}
    \caption{VLSTM PnL per asset - FX Futures} 
    \label{fig:vlstm_asset_fx}
\end{figure*}

\begin{figure}[h]
    \centering
    \includegraphics[width=\columnwidth]{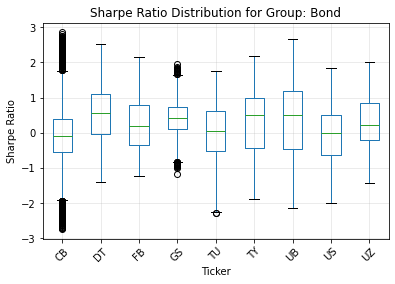}
    \caption{VLSTM box and whisker annual Sharpe Ratio per asset - Bond Futures}
    \label{fig:vlstm_asset_box_bond}
\end{figure}

\begin{figure}[t]
    \centering
    \includegraphics[width=\columnwidth]{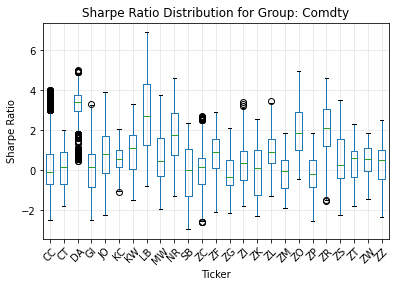}
    \caption{VLSTM box and whisker annual Sharpe Ratio per asset - Commodities Futures}
    \label{fig:vlstm_asset_box_cmdty}
\end{figure}

\begin{figure}[h]
    \centering
    \includegraphics[width=\columnwidth]{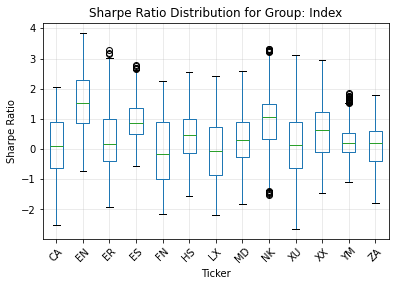}
    \caption{VLSTM box and whisker annual Sharpe Ratio per asset - Index Futures}
    \label{fig:vlstm_asset_box_index}
\end{figure}

\begin{figure}[t]
    \centering
    \includegraphics[width=\columnwidth]{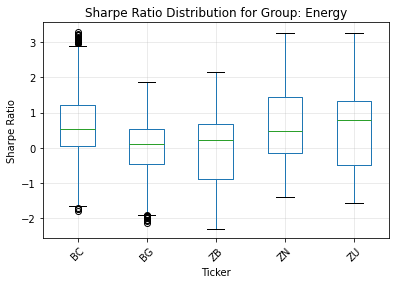}
    \caption{VLSTM box and whisker annual Sharpe Ratio per asset - Energy Futures}
    \label{fig:vlstm_asset_box_energy}
\end{figure}

\begin{figure}[h]
    \centering
    \includegraphics[width=\columnwidth]{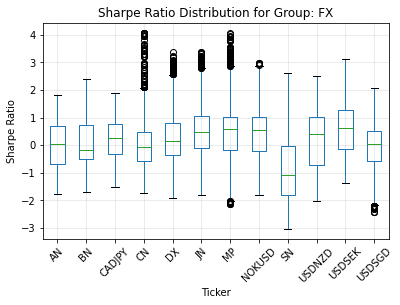}
    \caption{VLSTM box and whisker annual Sharpe Ratio per asset - FX Futures}
    \label{fig:vlstm_asset_box_fx}
\end{figure}

Figures~\ref{fig:vlstm_asset_bond}--\ref{fig:vlstm_asset_fx} display cumulative PnL per asset across the five asset categories. The corresponding box-and-whisker plots (Figures~\ref{fig:vlstm_asset_box_bond}--\ref{fig:vlstm_asset_box_fx}) summarize the distribution of Sharpe Ratio for each instrument.

Several observations emerge. First, performance is not driven by a single dominant contract but is distributed across multiple assets within each category. Second, tail behavior, as captured in the box plots, suggests that risk-adjusted performance is primarily driven by stable median returns rather than isolated extreme gains.

Overall, the per-asset analysis indicates that VLSTM’s aggregate performance is supported by consistent cross-sectional contributions rather than concentration in a small subset of instruments.

\FloatBarrier
\begin{table*}[t]
\centering
\scriptsize
\setlength{\tabcolsep}{3pt}
\caption{VLSTM: Annualised Volatility-Rescaled Performance and Breakeven Transaction Costs (bps)}
\begin{tabular}{llll | llll}
\hline
Ticker & Gross (ann.) & Turnover (ann.) & $c^*$ (bps) &
Ticker & Gross (ann.) & Turnover (ann.) & $c^*$ (bps) \\
\hline
LB & 0.2485 & 80.58 & 30.84 &
ZO & 0.2027 & 84.01 & 24.13 \\
DA & 0.2529 & 114.78 & 22.04 &
ZN & 0.0618 & 45.04 & 13.73 \\
JO & 0.0986 & 74.19 & 13.29 &
EN & 0.1003 & 86.69 & 11.57 \\
KC & 0.0738 & 63.92 & 11.54 &
ZR & 0.1339 & 137.44 & 9.75 \\
BC & 0.0512 & 61.26 & 8.36 &
ZF & 0.1279 & 157.12 & 8.14 \\
NR & 0.1086 & 137.78 & 7.88 &
NK & 0.0706 & 89.77 & 7.87 \\
ES & 0.0796 & 107.12 & 7.43 &
KW & 0.0686 & 93.77 & 7.32 \\
ZL & 0.0827 & 115.79 & 7.15 &
ZA & 0.0558 & 82.87 & 6.74 \\
ZI & 0.0444 & 68.93 & 6.44 &
ZU & 0.0381 & 64.42 & 5.91 \\
ZS & 0.0476 & 83.69 & 5.68 &
MW & 0.0669 & 127.30 & 5.25 \\
XU & 0.0507 & 98.17 & 5.17 &
XX & 0.0614 & 118.76 & 5.17 \\
ZK & 0.0421 & 83.07 & 5.06 &
CT & 0.0421 & 91.73 & 4.59 \\
ZW & 0.0417 & 91.08 & 4.57 &
SB & 0.0313 & 80.33 & 3.89 \\
MD & 0.0442 & 115.53 & 3.82 &
HS & 0.0301 & 85.51 & 3.52 \\
ZC & 0.0363 & 103.71 & 3.50 &
ZT & 0.0464 & 135.72 & 3.42 \\
ER & 0.0264 & 80.13 & 3.30 &
CA & 0.0267 & 92.21 & 2.90 \\
CR & 0.0389 & 138.25 & 2.82 &
BG & 0.0216 & 78.91 & 2.73 \\
DT & 0.0782 & 291.86 & 2.68 &
MP & 0.0368 & 143.21 & 2.57 \\
ZM & 0.0215 & 83.55 & 2.57 &
GI & 0.0230 & 94.01 & 2.44 \\
ZB & 0.0160 & 66.47 & 2.41 &
NOKUSD & 0.0382 & 165.09 & 2.31 \\
USDSEK & 0.0419 & 188.00 & 2.23 &
GS & 0.0547 & 282.39 & 1.94 \\
JN & 0.0321 & 209.92 & 1.53 &
ZG & 0.0177 & 119.59 & 1.48 \\
CC & 0.0115 & 78.16 & 1.47 &
USDNZD & 0.0209 & 169.47 & 1.23 \\
USDSGD & 0.0422 & 368.09 & 1.15 &
UB & 0.0664 & 591.04 & 1.12 \\
ZP & 0.0110 & 101.90 & 1.08 &
TY & 0.0415 & 411.13 & 1.01 \\
DX & 0.0284 & 298.04 & 0.95 &
LX & 0.0101 & 109.94 & 0.91 \\
YM & 0.0085 & 106.28 & 0.80 &
CN & 0.0199 & 272.83 & 0.73 \\
CB & 0.0194 & 344.26 & 0.56 &
FN & 0.0130 & 233.13 & 0.56 \\
CADJPY & 0.0080 & 151.64 & 0.53 &
US & 0.0104 & 198.09 & 0.52 \\
FB & 0.0281 & 689.19 & 0.41 &
UZ & 0.0508 & 1940.20 & 0.26 \\
TU & 0.0094 & 2218.65 & 0.04 &
ZZ & -0.0004 & 80.57 & -0.05 \\
BN & -0.0092 & 207.76 & -0.44 &
AN & -0.0170 & 164.71 & -1.03 \\
SN & -0.0272 & 212.34 & -1.28 \\
\hline
\end{tabular}
\label{tbl:vlstm_tcost}
\end{table*}

\subsection{xLSTM}
\begin{figure*}[h]
    \centering
    \includegraphics[width=0.8\textwidth]{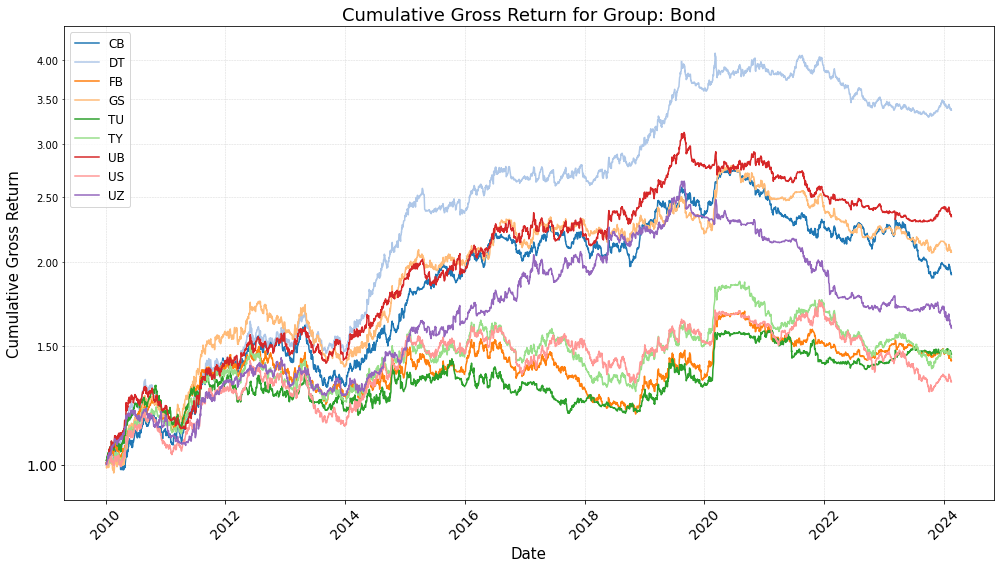}
    \caption{xLSTM PnL per asset - bond Futures} 
    \label{fig:xlstm_asset_bond}
\end{figure*}

\begin{figure*}[h]
    \centering
    \includegraphics[width=0.8\textwidth]{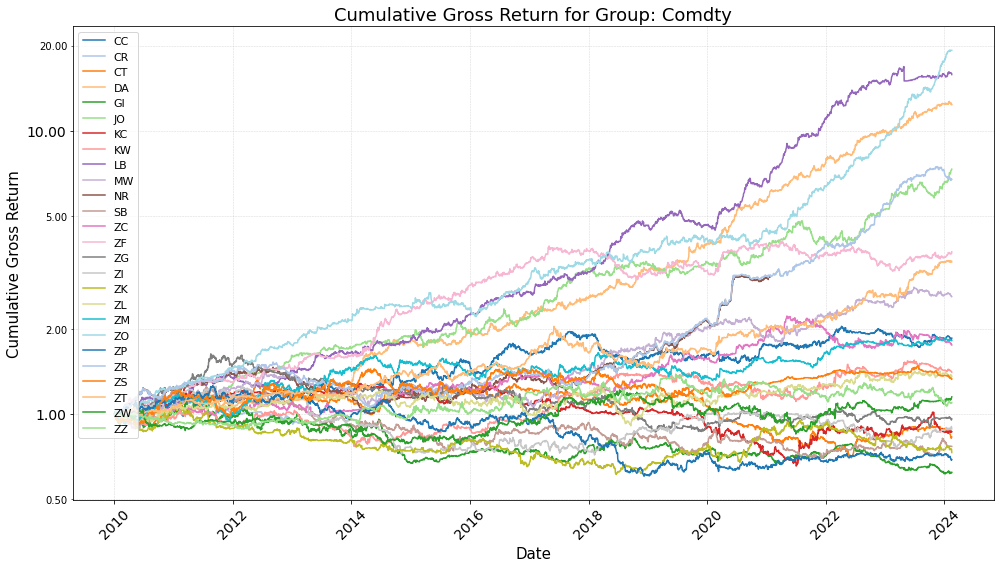}
    \caption{xLSTM PnL per asset - Commodities Futures}
    \label{fig:xlstm_asset_cmdty}
\end{figure*}

\begin{figure*}[h]
    \centering
    \includegraphics[width=0.8\textwidth]{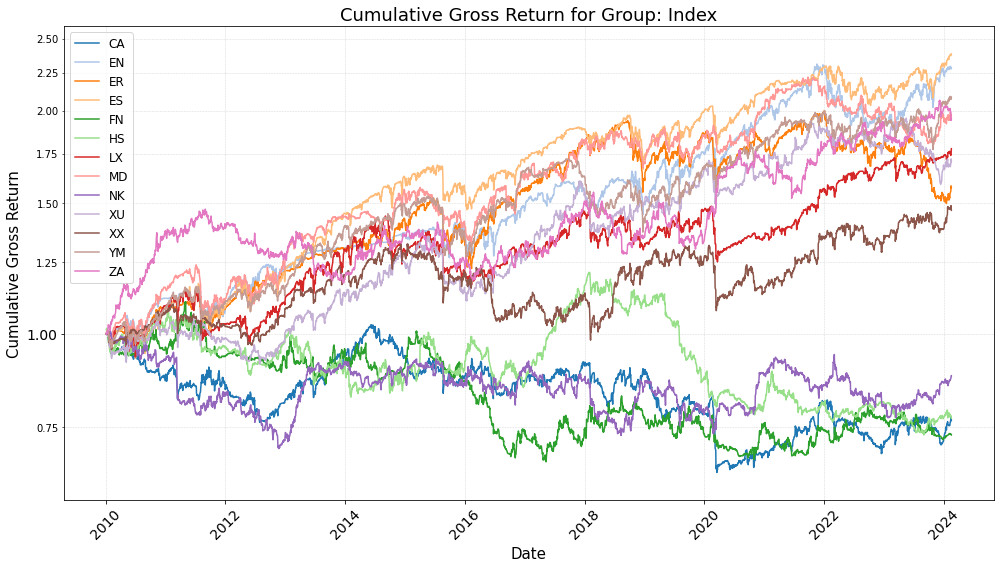}
    \caption{xLSTM PnL per asset - Index Futures}
    \label{fig:xlstm_asset_index}
\end{figure*}

\begin{figure*}[h]
    \centering
    \includegraphics[width=0.8\textwidth]{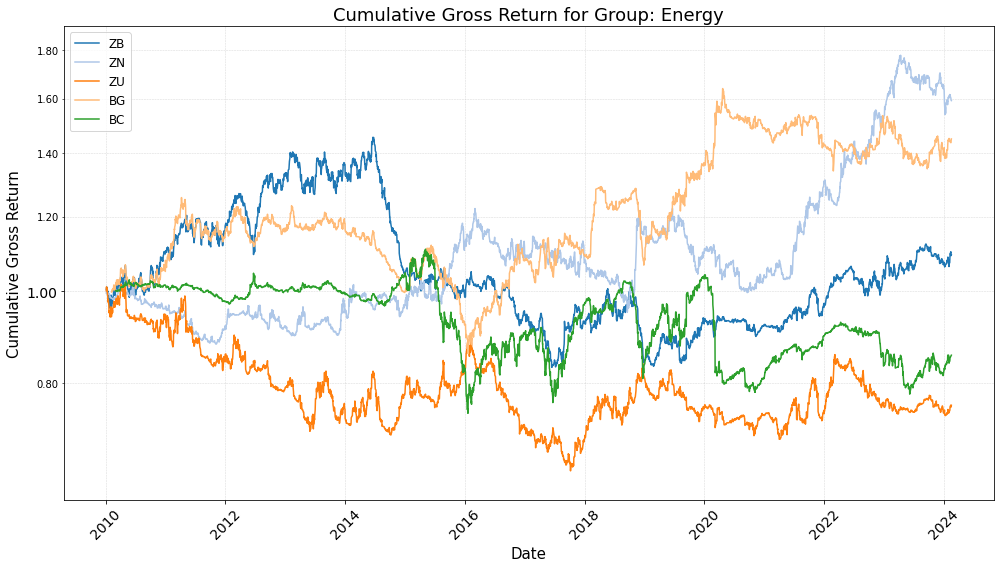}
    \caption{xLSTM PnL per asset - Energy Futures}
    \label{fig:xlstm_asset_energy}
\end{figure*}

\begin{figure*}[h]
    \centering
    \includegraphics[width=0.8\textwidth]{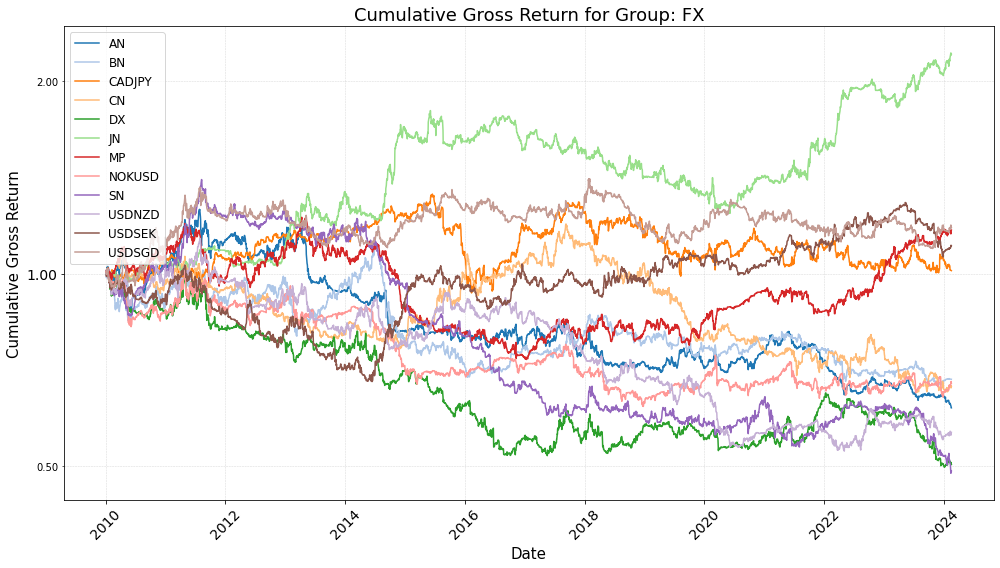}
    \caption{xLSTM PnL per asset - FX Futures}
    \label{fig:xlstm_asset_fx}
\end{figure*}

\begin{figure}[h]
    \centering
    \includegraphics[width=\columnwidth]{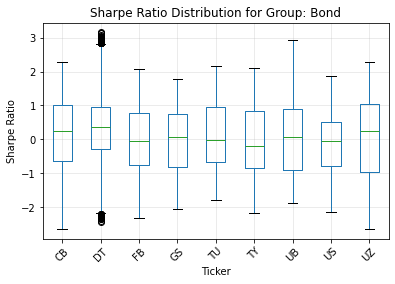}
    \caption{xLSTM box and whisker annual Sharpe Ratio per asset - Bond Futures 
    }
    \label{fig:xlstm_asset_box_bond}
\end{figure}

\begin{figure}[h]
    \centering
    \includegraphics[width=\columnwidth]{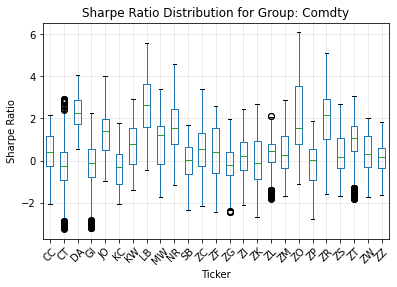}
    \caption{xLSTM box and whisker annual Sharpe Ratio per asset - Commodities Futures}
    \label{fig:xlstm_asset_box_cmdty}
\end{figure}

\begin{figure}[h]
    \centering
    \includegraphics[width=\columnwidth]{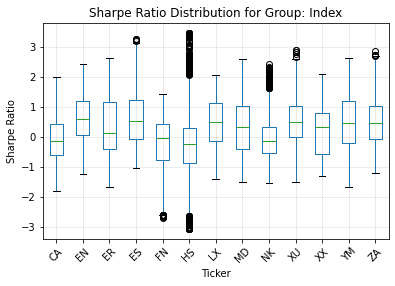}
    \caption{xLSTM box and whisker annual Sharpe Ratio per asset - Index Futures}
    \label{fig:xlstm_asset_box_index}
\end{figure}

\begin{figure}[h]
    \centering
    \includegraphics[width=\columnwidth]{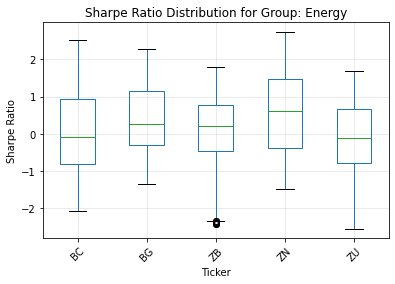}
    \caption{xLSTM box and whisker annual Sharpe Ratio per asset - Energy Futures}
    \label{fig:xlstm_asset_box_energy}
\end{figure}

\begin{figure}[h]
    \centering
    \includegraphics[width=\columnwidth]{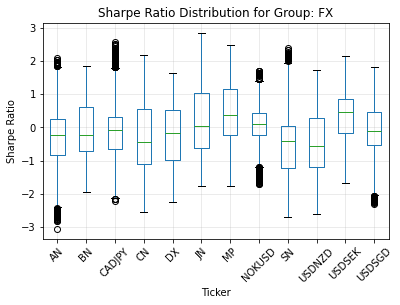}
    \caption{xLSTM box and whisker annual Sharpe Ratio per asset - FX Futures}
    \label{fig:xlstm_asset_box_fx}
\end{figure}

Figures~\ref{fig:xlstm_asset_bond}--\ref{fig:xlstm_asset_fx} present cumulative PnL per asset for xLSTM across the five asset groups. The associated box-and-whisker plots (Figures~\ref{fig:xlstm_asset_box_bond}--\ref{fig:xlstm_asset_box_fx}) provide distributional summaries.

Relative to VLSTM, xLSTM exhibits stronger cross-sectional homogeneity in certain categories. In higher-volatility sectors such as energy and commodities, dispersion increases, but median performance remains positive across most instruments. This suggests that the model adapts to heterogeneous volatility structures without excessive tail risk concentration.

Importantly, no systematic degradation is observed in a specific asset class, indicating that performance is not regime- or sector-dependent. The cross-asset consistency observed in the box plots further supports the robustness of the learned representations.

\FloatBarrier
\begin{table*}[t]
\centering
\scriptsize
\setlength{\tabcolsep}{3pt}
\caption{xLSTM: Annualised Volatility-Rescaled Performance and Breakeven Transaction Costs (bps)}
\begin{tabular}{llll | llll}
\hline
Ticker & Gross (ann.) & Turnover (ann.) & $c^*$ (bps) &
Ticker & Gross (ann.) & Turnover (ann.) & $c^*$ (bps) \\
\hline
LB & 0.2338 & 68.93 & 33.92 &
DA & 0.2200 & 83.29 & 26.41 \\
ZO & 0.2157 & 98.88 & 21.81 &
ES & 0.0501 & 26.01 & 19.25 \\
ZN & 0.0544 & 29.02 & 18.76 &
CC & 0.0419 & 24.23 & 17.31 \\
EN & 0.0572 & 38.19 & 14.97 &
JO & 0.1375 & 96.18 & 14.30 \\
ZR & 0.1818 & 132.86 & 13.68 &
NR & 0.1489 & 135.22 & 11.01 \\
LX & 0.0430 & 40.14 & 10.72 &
BG & 0.0443 & 51.13 & 8.66 \\
ZM & 0.0279 & 32.54 & 8.57 &
KW & 0.0587 & 75.50 & 7.77 \\
YM & 0.0409 & 55.80 & 7.33 &
MW & 0.0824 & 116.51 & 7.07 \\
ZT & 0.0919 & 131.96 & 6.96 &
MP & 0.0419 & 61.25 & 6.83 \\
DT & 0.0501 & 79.15 & 6.33 &
ZA & 0.0437 & 76.56 & 5.71 \\
ZW & 0.0415 & 84.22 & 4.93 &
ZC & 0.0471 & 101.33 & 4.65 \\
MD & 0.0325 & 71.83 & 4.53 &
JN & 0.0336 & 74.70 & 4.50 \\
ZZ & 0.0167 & 42.08 & 3.96 &
ZF & 0.0580 & 161.56 & 3.59 \\
ZI & 0.0211 & 60.24 & 3.50 &
ZS & 0.0121 & 38.73 & 3.12 \\
ER & 0.0163 & 52.64 & 3.09 &
CR & 0.0389 & 138.25 & 2.89 \\
USDSEK & 0.0286 & 100.11 & 2.85 &
ZL & 0.0314 & 118.88 & 2.64 \\
ZB & 0.0107 & 41.06 & 2.62 &
XU & 0.0385 & 151.47 & 2.54 \\
UB & 0.0305 & 176.77 & 1.72 &
XX & 0.0192 & 178.70 & 1.07 \\
CB & 0.0177 & 217.61 & 0.81 &
GS & 0.0133 & 223.73 & 0.60 \\
NK & 0.0055 & 145.81 & 0.38 &
ZU & 0.0011 & 38.32 & 0.29 \\
ZK & 0.0032 & 115.80 & 0.28 &
UZ & 0.0145 & 584.56 & 0.25 \\
FB & 0.0087 & 441.46 & 0.20 &
TU & 0.0205 & 1121.50 & 0.18 \\
TY & 0.0034 & 219.72 & 0.16 &
USDSGD & 0.0010 & 208.81 & 0.05 \\
NOKUSD & -0.0028 & 102.23 & -0.27 &
CN & -0.0106 & 193.52 & -0.55 \\
US & -0.0086 & 101.15 & -0.85 &
CADJPY & -0.0144 & 166.63 & -0.86 \\
GI & -0.0072 & 54.74 & -1.31 &
DX & -0.0359 & 270.55 & -1.33 \\
BN & -0.0143 & 95.74 & -1.49 &
CA & -0.0167 & 110.31 & -1.52 \\
SB & -0.0084 & 41.65 & -2.01 &
HS & -0.0103 & 45.01 & -2.30 \\
FN & -0.0193 & 77.30 & -2.50 &
AN & -0.0371 & 131.94 & -2.81 \\
ZP & -0.0301 & 100.81 & -2.99 &
ZG & -0.0190 & 63.04 & -3.02 \\
USDNZD & -0.0386 & 92.40 & -4.17 &
CT & -0.0397 & 92.56 & -4.29 \\
KC & -0.0218 & 43.61 & -4.99 &
SN & -0.0737 & 125.71 & -5.86 \\
BC & -0.0126 & 19.67 & -6.40 \\
\hline
\end{tabular}
\label{tbl:xlstm_tcost}
\end{table*}

\paragraph{Cross-Sectional Robustness.}

Taken together, the complementary results reinforce the conclusions drawn in the main text. Aggregate Sharpe ratios are not driven by isolated outliers but reflect broadly distributed performance across asset classes. Differences between models are manifested not only in overall portfolio metrics but also in dispersion characteristics and category-level stability.

This granular evaluation is particularly relevant in financial forecasting, where apparent portfolio-level improvements may otherwise mask instability or concentration risk at the instrument level.


\section{Annual Sharpe Ratio}
Table~\ref{tab:performance_by_year_extended} reports annual out-of-sample Sharpe ratios for all benchmark strategies over the period 2010--2024. The year-by-year decomposition complements the aggregated results in the main text by providing a finer assessment of temporal stability and regime dependence. Several patterns emerge. First, performance is not concentrated in a single favorable subperiod: the leading deep sequence models exhibit consistently positive Sharpe ratios across a broad range of market environments, including the post-crisis recovery, the low-volatility mid-2010s expansion, and the high-uncertainty period following 2020. Second, while cross-sectional performance rankings vary from year to year---as expected given structural shifts in volatility and cross-asset correlations---the top-performing architectures remain competitive in most years and avoid persistent underperformance. Third, classical linear benchmarks display greater sensitivity to adverse regimes, with more frequent negative annual Sharpe ratios. Overall, the annual breakdown confirms that the superior aggregated performance documented in the main text is not driven by isolated episodes, but rather reflects sustained risk-adjusted returns across heterogeneous market conditions.
\begin{table*}[t]
\centering
\caption{Annual Sharpe Ratios by Strategy (2010--2024)}
\resizebox{\textwidth}{!}{
\begin{tabular}{lrrrrrrrrrrrrrrr}
\toprule
Strategy & 2010 & 2011 & 2012 & 2013 & 2014 & 2015 & 2016 & 2017 & 2018 & 2019 & 2020 & 2021 & 2022 & 2023 & 2024 \\
\midrule
AR1x        & 1.45 & -0.07 & -0.11 & 0.23 & 2.97 & -0.02 & 0.05 & 0.67 & -0.59 & 0.18 & 3.02 & 0.63 & 1.37 & 1.45 & 0.30 \\
AR$n$x      & 1.60 & 0.06 & -0.25 & 0.06 & 2.51 & 0.22 & -0.17 & 0.30 & -0.90 & 0.48 & 2.07 & 0.90 & 1.64 & 1.11 & -0.16 \\
DLinear     & 1.21 & -0.11 & -0.43 & -0.29 & 2.89 & 0.33 & -0.19 & 0.42 & -0.28 & -0.27 & 3.22 & 0.35 & 1.45 & 2.30 & -0.94 \\
NLinear     & 1.25 & 0.25 & -0.85 & -0.18 & 2.58 & 0.10 & -0.08 & 1.13 & -0.94 & 0.77 & 2.74 & 1.29 & 0.12 & 0.87 & -1.19 \\
LSTM        & 3.58 & 0.29 & 2.47 & 0.74 & 1.81 & \textbf{2.09} & 2.19 & -1.19 & \textbf{1.72} & 3.20 & -1.51 & 1.12 & 3.44 & 1.22 & \underline{1.06} \\
VLSTM       & 3.96 & \underline{1.48} & \textbf{3.92} & \underline{1.13} & 3.03 & \underline{1.90} & \textbf{4.03} & 1.06 & 1.33 & \textbf{4.74} & 0.10 & 2.81 & \underline{3.82} & \underline{2.76} & -0.10 \\
VxLSTM      & 2.73 & 1.34 & 1.71 & 0.22 & 3.25 & 0.11 & 1.92 & \underline{2.82} & 0.60 & 1.95 & \underline{3.77} & 2.24 & 2.13 & 1.88 & -1.31 \\
xLSTM       & 2.64 & 1.41 & 1.72 & -0.28 & 2.91 & 0.34 & 2.29 & \textbf{2.96} & 0.46 & 2.38 & \textbf{3.83} & 2.62 & 1.55 & 1.96 & 0.02 \\
Mamba       & 0.82 & -0.27 & -0.46 & -0.48 & 3.03 & 0.41 & 0.07 & 0.04 & -0.17 & -0.40 & 1.29 & 1.00 & -0.54 & 1.47 & -0.40 \\
Mamba2      & 0.88 & -0.38 & -0.37 & 0.07 & 3.22 & 0.18 & 0.15 & 0.68 & -0.05 & 0.00 & 3.15 & 1.33 & 2.27 & 1.43 & -0.49 \\
VSN+Mamba2  & 1.48 & 0.27 & -0.12 & -0.00 & \textbf{3.43} & 0.65 & 0.54 & 1.31 & -0.05 & 0.23 & 3.31 & 1.92 & 2.64 & 1.45 & -0.63 \\
PatchTST    & 2.76 & 0.56 & 0.54 & -1.19 & 0.74 & 0.37 & 0.36 & -0.23 & 0.85 & 1.21 & 1.36 & 2.02 & 0.68 & 0.55 & 0.72 \\
LPatchTST   & \textbf{3.99} & \textbf{1.61} & 3.10 & 0.59 & \underline{3.32} & 1.67 & 1.88 & 2.09 & \underline{1.56} & 3.08 & 1.17 & \textbf{3.21} & 3.19 & 2.76 & \textbf{1.17} \\
PsLSTM      & 2.75 & 1.29 & 1.80 & -0.21 & 3.21 & 1.16 & 2.07 & 1.91 & 1.40 & 2.50 & 1.95 & \underline{3.00} & 1.92 & \underline{2.79} & -0.89 \\
TFT         & \underline{3.97} & 1.21 & \underline{3.90} & \textbf{1.19} & 3.04 & 1.50 & \underline{3.18} & 1.05 & 1.32 & \underline{3.36} & -0.14 & 2.62 & \textbf{4.28} & \textbf{3.25} & 0.37 \\
iTransformer& 1.30 & 0.20 & -1.14 & 0.19 & 2.34 & 0.70 & -0.07 & 0.65 & -1.16 & 0.21 & 1.43 & 0.95 & -1.03 & 0.68 & 0.46 \\
\bottomrule
\end{tabular}
}
\label{tab:performance_by_year_extended}
\end{table*}